\documentclass[fleqn,usenatbib]{mnras}

\usepackage{newtxtext,newtxmath,appendix}

\usepackage[T1]{fontenc}

\DeclareRobustCommand{\VAN}[3]{#2}
\let\VANthebibliography\thebibliography
\def\thebibliography{\DeclareRobustCommand{\VAN}[3]{##3}\VANthebibliography}

\usepackage{graphicx}	
\usepackage{amsmath}	
\usepackage{xcolor}

\providecommand{\teff}{$T_{\rm eff}$}
\providecommand{\feh}{[Fe/H]}
\providecommand{\logg}{$\log g$}

\newcommand{\Cannon}{{\it The Cannon}}

\newcommand{\highSNRsample}{$5\,040$}

\title[Phylogenetics in GALAH]{Assembling a high-precision abundance catalogue of solar twins in GALAH for phylogenetic studies}

\author[Walsen et al.]{
Kurt Walsen,$^{1, 2, 3}$\thanks{E-mail: kurt.walsen.b@gmail.com}
Paula Jofr\'e,$^{2, 3}$
Sven Buder,$^{4,5}$
Keaghan Yaxley,$^{6}$
Payel Das,$^{7}$
Robert Yates,$^{8}$
\newauthor
Xia Hua,$^{9}$
Theosamuele Signor,$^{3,10}$
Camilla Eldridge,$^{2, 3}$
Alvaro Rojas-Arriagada$^{2, 11}$,
Patricia Tissera$^{2,12, 13}$, 
\newauthor
Evelyn Johnston$^{2,3}$,
Claudia Aguilera-G\'omez$^{12}$, Manuela Zoccali$^{12,14}$, Gerry Gilmore$^{15, 16}$, Robert Foley,$^{17}$  \\
$^{1}$Departamento de Ingenier\'ia Matem\'atica, Facultad de Ciencias F\'isicas y Matem\'aticas, Universidad de Chile, Av. Beauchef 851, Santiago, Chile \\
$^{2}$Millenium Nucleus ERIS \\
$^{3}$Instituto de Estudios Astrof\'isicos, Facultad de Ingenier\'ia y Ciencias, Univesidad Diego Portales, Av. Ej\'ercito 441, Santiago, Chile \\
$^{4}$Research School of Astronomy \& Astrophysics, Australian National University, Canberra, ACT 2611, Australia\\
$^{5}$ARC Centre of Excellence for All Sky Astrophysics in 3 Dimensions (ASTRO 3D), Australia\\
$^{6}$Research School of Biology, Australian National University, Canberra , ACT 2601, Australia\\
$^{7}$Physics Department, University of Surrey, Guildford GU2 7XH, United Kingdom\\
$^{8}$Centre for Astrophysics Research, University of Hertfordshire, Hatfield, AL10 9AB, United Kingdom\\
$^{9}$Mathematical Sciences Institute, Australian National University, Canberra ACT 0200, Australia\\
$^{10}$Inria Chile Research Center, Av. Apoquindo 2827, piso 12, Santiago, Chile\\
$^{11}$Departamento de F\'isica, Universidad de Santiago de Chile, Av. Victor Jara 3659, Santiago, Chile\\
$^{12}$Instituto de Astrof\'isica, Pontificia Universidad Cat\'olica de Chile, Av. Vicuña Mackenna 4860, Santiago, Chile \\
$^{13}$Centro de Astro-Ingenier\'ia, Pontificia Universidad Cat\'olica de Chile, Av. Vicuña Mackenna 4860, Santiago, Chile \\
$^{14}$Millennium Institute of Astrophysics, Av. Vicuña Mackenna 4860, 82-0436 Macul, Santiago, Chile \\
$^{15}$Institute of Astronomy, University of Cambridge, Madingley Road, Cambridge CB3 0HA, UK\\
$^{16}$ Institute of Astrophysics, FORTH Cert, N. Plastira 100 Vassilika, Crete, Greece\\
$^{17}$Leverhulme Centre for Human Evolutionary Studies, Department for Anthropology and Archaeology, University of Cambridge, CB2 1QH, UK\\
}
\date{Accepted XXX. Received YYY; in original form ZZZ}

\pubyear{2015}

\begin{document}
\label{firstpage}
\pagerange{\pageref{firstpage}--\pageref{lastpage}}
\maketitle

\begin{abstract}
Stellar chemical abundances have proved themselves a key source of information for understanding the evolution of the Milky Way, and the scale of major stellar surveys such as GALAH have massively increased the amount of chemical data available. However, progress is hampered by the level of precision in chemical abundance data { as well as the visualization methods for comparing the multidimensional outputs of} chemical evolution {models} to stellar abundance data. Machine learning  methods have greatly improved the former; while the application of tree-building or phylogenetic methods borrowed from biology are beginning to show promise with the latter. Here we analyse a sample of GALAH solar twins to address these issues. We apply \Cannon\ algorithm \citep{ness-15} to generate a catalogue of about 40,000 solar twins with 14 high precision abundances which we use to perform a phylogenetic analysis on a selection of stars that have two different ranges of eccentricities. From our analyses we are able to find a group with mostly stars on circular orbits and some old stars with eccentric orbits whose age-[Y/Mg] relation agrees remarkably well with the chemical clocks published by previous high precision abundance studies. Our results show the power of combining survey data with machine learning and phylogenetics to reconstruct the history of the Milky Way.

\end{abstract}




\section{Introduction}

Chemical enrichment plays an important role in the formation and evolution of our Galaxy. Significant advances in our understanding of galaxy evolution have come from interpreting data from major stellar spectroscopic surveys, such as the Galactic Archaeology with HERMES (GALAH) Survey  \citep{DeSilva2015, GALAH-DR3}, the Gaia-ESO Survey \citep{gilmore-23, randich-22} and the  Sloan Digital Sky Survey \citep{sdss-dr17}. These surveys have observed hundreds of thousands of stars for which chemical abundances of up to 30 elements have been measured  \citep[see extensive discussion in][]{jofre-19}. However, while this increase of data is essential, it is also important to develop tools for analysing larger datasets. In this paper we are concerned with the presentation and assessment of such methods in two ways – first, the extraction of chemical information from stars via stellar spectroscopy, and secondly, the analysis of chemical abundances to retrieve an evolutionary history of Galactic chemical enrichment. 

Galactic history shares some features with biological evolution. Indeed, \cite{Jofre-17Phylo} and \cite{Jackson-21} showed that phylogenetic techniques can be employed to reconstruct the evolution of star formation within the solar neighbourhood using a small but very precise sample of stellar abundances of solar twins. Interpreting their results was difficult because of selection effects in their stellar samples, which were a rather small sample of spectra obtained from public archives \citep{Nissen-16, Bedell-18}.  While major spectroscopic surveys retrieve data on millions of stars observed with a well-defined selection function, there is a trade-off between the scale of data and its resolution. This study is thus motivated by carefully creating a sample of high-precision chemical abundances from survey data. We then use a phylogenetic tree algorithm to assess the formation sequences of the Milky Way. First, though, we discuss the potential and dangers of using machine learning tools (ML) to extract higher precision chemical abundances from stellar spectra for this purpose and then we introduce phylogenetic methods.

\subsection{Chemical abundances from stellar spectra}

In this era of large spectroscopic surveys, ML has become a revolutionary way to both, precisely and quickly derive spectral properties \citep[see e.g.][to name a few]{Ambrosch-23, Wheeler-20, LeungBovy}.  The revolution became notable perhaps with the introduction of  {\it The Cannon} \citep{ness-15} into the field.  The method finds a polynomial function that directly relates the spectra with labels. To do so, the function is found from stellar spectra for which these labels are known. {\it The Cannon} is very fast, and provides more precise results than standard methods when the spectra are noisy or not of very high resolution. Since it is easy to implement, it has quickly been applied on a large variety of stellar spectra \citep{Casey-17, galah-dr2, Wheeler-20, nandakumar-22}.  Many other spectral analyses using ML have followed \citep{Ting-Payne,  LeungBovy, Guiglion-20}.  Nowadays there is a large variety of chemical data products derived from neural networks or mathematical functions trained on synthetic or observed spectral grids. Since ML allows labels to be transferred from one survey to the other \citep{Wheeler-20, nandakumar-22}, it is possible to put several surveys on the same scale provided there are stars in common between surveys \citep{Ness-data-driven}. Indeed, ML is so powerful that today it might seem absurd to aim performing a spectroscopic analysis on million spectra with the ``standard" methods \citep[see discussion of such methods in][]{jofre-19}.

Although methods like \Cannon\ provide many advantages over the standard methods, it relies on training sets that are calibrated with the standard analyses. Choosing the training set is not straightforward, it should sample fully and evenly the parameter space of the test set, and should have accurate and precise values for the labels that want to be determined. A grid of synthetic spectra is powerful because it ensures even sampling \citep{Ting-Payne}, but synthetic spectra and real spectra are different from each other leading to a level of uncertainty in the results \citep[see discussion in][]{cycle-net}. ML methods are suitable when the methods used on the observed test stars are the same as for the training set of stars, especially if there is information additional to the spectra (interferometry, accurate parallaxes, astroseismology, etc). This extra information provides higher accuracy for the labels \citep{Miglio13,JofreGBS, Heiter15}. The problem here, though, is the sampling.  ML works better with large datasets, and so using the largest number of stars as possible is preferable. However, the full parameter space is not often available in these very large datasets, and so there is a trade-off between sample size and data completeness.  It is thus  worth investigating systematically how to train a ML algorithm with the more limited data available today.

\subsection{Phylogenetic trees as a promising tool to trace Galactic chemical evolution} \label{sec:intro_phylo}

Phylogenetic trees are graphs that illustrate the shared evolutionary history among a dataset, allowing us to understand the hierarchical pattern of ancestry and descent which connects all of the observations \citep{Baum2005}. Phylogenetic methods can reconstruct ancestral relationships as long as there is a shared history and a heritable process linking the data objects. These objects are normally individuals, species and higher taxa in biology, where methods to analyse them have been developed \citep{felsenstein_molecular}, but they are applicable more broadly. By making the hypothesis that the stars in the Milky Way disk come from the same but evolving interstellar medium (ISM), and that the evolutionary marker (i.e. the heritable component) of the ISM is the chemical composition, we can use the chemical abundances of low-mass stars as fossil records for building phylogenetic trees \citep[see also][]{Freeman2002}.

The hypothesis that stars in the Milky Way form from the same evolving ISM is a simplification of reality. Indeed, the Milky Way has accreted other dwarf galaxies, depositing in the ISM some gas that has been enriched by a different chemical evolutionary history. An example is the interaction of the Milky Way with Sagittarius, which has affected the star formation history of the disk significantly \citep{Ruiz-lara20}. This is not a limitation for phylogenetic trees however, ``hybridization" processes are common in biology, and ways to characterizing their impact are active topics of research. Another simplification of this hypothesis is the fact that chemical abundances in low-mass stars are not as constant over their lifetime as we would like them to be. Heavy elements sink in the atmospheres due to gravitational settling \citep{Lind08, Souto18}, causing an effect in the measured abundances depending on the age and the mass. This issue has an impact on chemical tagging studies overall. 

Phylogenetic trees have already been constructed in \cite{Jofre-17Phylo} and \cite{Jackson-21}. These papers focused on solar twins for the practical reason that estimates of chemical abundances in solar twins are very precise, particularly if they are derived differentially with respect to the Sun \citep[e.g.][]{Nissen-18Rev}. \cite{Jofre-17Phylo} used high precision data published by \cite{Nissen2015, Nissen-16} and \cite{Jackson-21} the data published by \cite{Bedell-18}. The trees were built using a nearest neighbourhood distance method, which essentially considers the pairwise distance in chemical abundances between stars to find the hierarchical differences, displaying them in a tree. \cite{Jofre-17Phylo} found a tree with different branches where the relationship between branch length and age was different, suggesting that with trees it might be possible to identify stellar families (i.e. groups of stars that cluster together and so may have a shared history), but more importantly, study their different evolutionary processes such as chemical enrichment rate. \cite{Jackson-21} followed that study by enlarging the number of stars and by choosing elemental abundance ratios which evolve with time, e.g. the so-called {\it chemical clocks} \citep[e.g.,][]{Nissen-16, Casali-2020, Jofre-20}. They also found different branches that had different ages and dynamical distributions, and attributed them to different stellar groups co-existing in the solar neighbourhood. How far these stellar groups were representative of the broader stellar population remains uncertain due to the selection function. 

\subsection{Aim and structure of this study} \label{sec:intro_structure}

In this work  we make the first study of a phylogenetic tree from survey data, for which we chose a set of solar twins from GALAH~DR3 \citep{GALAH-DR3}. We compare our tree using the published abundances from GALAH and a set of abundances obtained using ML, for which the precision is higher. To do so, we first apply the spectral fitting machinery of  \Cannon\ to  GALAH data \citep{GALAH-DR3} to improve the precision of the chemical abundance measurements. We systematically assess the steps in training \Cannon\ in solar twins.  

GALAH has observed a very large sample of solar twins (to date about 40,000).  By carefully applying \Cannon\ we provide a sample of high precision abundances for a much larger sample of solar twins that those previously published from high resolution data. This takes the sample size from around 500 stars \citep{Casali-2020} to two orders of magnitude. We then select a sample of this catalogue to test if the phylogenetic signal improves from GALAH to \Cannon\ abundances for the same stars. 

The paper is structured as follows: In the next section we describe the data used, and this is followed by a description of how  \Cannon\ training is set up. Our new catalogue is presented in Sec.~\ref{sec:catalogue} and we apply phylogenetic techniques to analyse this catalogue in Sec.~\ref{sec:applications}. Some general conclusions are presented in Sec.~\ref{sec:conclusion}.

\section{Data and methods} \label{sec:data}

In this work we use stellar spectra, parameters, and abundances published as part of the third data release of GALAH \citep[][hereafter GALAH~DR3]{GALAH-DR3}. Stellar spectra were observed with the HERMES spectrograph \citep{Sheinis2015} at the Anglo-Australian Telescope with a resolution of $\sim 28\,000$ across four wavelength regions in the optical (see Tab.~\ref{tab:wavelength_grid}) that include absorption features of more than 30 elements. For this study, we use the radial velocity corrected and normalised spectra downloaded from the \textsc{datacentral}\footnote{\url{https://datacentral.org.au/services/download/}} web interface and interpolate them onto a common wavelength scale given in Tab.~\ref{tab:wavelength_grid}. Flux uncertainties are derived from the provided relative error spectra and multiplied with the flux. 

We use the stellar parameters ($T_\text{eff}$, $\log g$, and [Fe/H]) as well as logarithmic elemental abundances relative to the Sun and iron, that is, [X/Fe]. These were extracted from the GALAH spectra via $\chi^2$ optimisation of on-the-fly computed synthetic spectra with Spectroscopy Made Easy \citep[\textsc{sme}][]{Valenti1996, Piskunov2017}. Synthetic spectra were computed based on 1D \textsc{marcs} model atmospheres \citep{Gustafsson2008} and non-local thermodynamic equilibrium for eleven elements \citep{Amarsi2020} and local thermodynamic equilibrium for all other elements. The optimisation included the constraint of surface gravities $\log g$ from bolometric luminosity estimates based on photometric information from the 2MASS survey \citep{Skrutskie2006} and distances inferred from the \textit{Gaia} satellite's second data release \citep{Brown2018, BailerJones2018}. 

The GALAH~DR3 catalogue further provides age estimates, which are needed for evolutionary studies. These ages and their uncertainties are estimated via isochrone interpolation with stellar parameters through the Bayesian fitting machinery \texttt{BSTEP} \citep{Sharma2018}. Valued-added catalogues in GALAH also include information about orbital properties, such as total velocities, eccentricities and actions. For more details we refer the reader to \citet{GALAH-DR3}.

\begin{table}
    \centering
    \begin{tabular}{ccccc}
    \hline
    CCD & Begin / $\text{\AA}$ & End / $\text{\AA}$ & Dispersion / $\text{\AA/pixel}$ & Nr. pixels \\
    \hline
    1 &	4715.94 & 4896.00 &	0.046 & 3915\\
	2 &	5650.06 & 5868.25 & 0.055 & 3968\\
	3 &	6480.52 & 6733.92 & 0.064 & 3960\\
	4 &	7693.50 & 7875.55 & 0.074 & 2461 \\
 \hline
    \end{tabular}
    \caption{Wavelength grid. Selected from GALAH-DR2 \citep{galah-dr2}.}
    \label{tab:wavelength_grid}
\end{table}

\subsection{Abundance uncertainties}
The total error budget in GALAH considers a combination of precision and accuracy uncertainties. Precision uncertainties are calculated from the internal covariance uncertainties of \textsc{sme}, that is, the uncertainties are computed from the diagonal of the covariance matrix given by \textsc{sme} fitting procedures, which were adjusted to be consistent with the scatter of repeat observations as a function of the signal-to-noise ratio (SNR). For this work, we consider the precision to be the maximum between both uncertainties described above. Accuracy uncertainties for the stellar parameters are derived from comparisons with reference stars, such as the Gaia FGK benchmark stars \citep{JofreGBS, Heiter15, Jofre2018}, whereas for the abundances we only have precision uncertainties reported. We thus consider for this work the GALAH~DR3 precision uncertainties, which only report the final abundance uncertainty based on the maximum of the internal covariance error for the abundance fit and the SNR response to repeat observations \citep{GALAH-DR3}.

\subsection{Solar twin selection}

We are interested in the solar twins of GALAH~DR3. We select them from the file called  \texttt{GALAH\_DR3\_main\_allspec\_v2.fits} \footnote{\url{https://www.galah-survey.org/dr3/the_catalogues/}}. This file contains  measurements of astrophysical parameters and chemical abundances of 678\,423 spectra from 588\,571 stars derived as explained above. We consider only spectra that have a signal-to-noise ratio (SNR) above 10 in all 4 CCDs.  In order to select the solar twins from that sample,  we used the reported astrophysical parameters and performed the following cuts:

\begin{align}
    | \mathrm{T}_{\mathrm{eff}} - 5\,777 | &< 200\  \mathrm{K} \nonumber\\
    | \log g - 4.44 | &< 0.3  \label{eq:sp_cuts}\\
    | \mathrm{[Fe/H]} | &< 0.3 \nonumber
\end{align}
 where we assume that the solar temperature is $5\,777~\mathrm{K}$, the solar surface gravity is $4.44$ and the logarithmic number density of iron to hydrogen compared to the sun, [Fe/H], is $0.0$ \citep{Prsa2016}. 

We further used the quality flags on all the chemical abundances available in the dataset, specifically \texttt{flag\_sp} and \texttt{flag\_X\_fe}, to verify which stars had reliable astrophysical parameters.   When the flag is zero it means that the measurement of the stellar parameters and chemical abundances have no reported problem. We also considered a high detection rate of the different elements, i.e. stars with several elements measured. Only 14 elements have a detection rate above 90\% in solar twins.  These elements correspond to Na (100\%), Mg (100\%), Al (98\%), Si (100\%), Ca (97\%), Sc (100\%), Ti (97\%), Cr (100\%), Mn (100\%), Ni (94\%), Cu (98\%), Zn (99\%), Y (94\%) and Ba (100\%).  Finally, we removed stars that had reported chemical abundance ratios with $\mathrm{[X/Fe]} > 1$ or $\mathrm{[X/Fe]} < -1$ to avoid  outliers.  All these cuts gave us 44\,317 entries, which correspond to 39\,554 spectra. 

\begin{figure*}
    \includegraphics[scale=0.38]{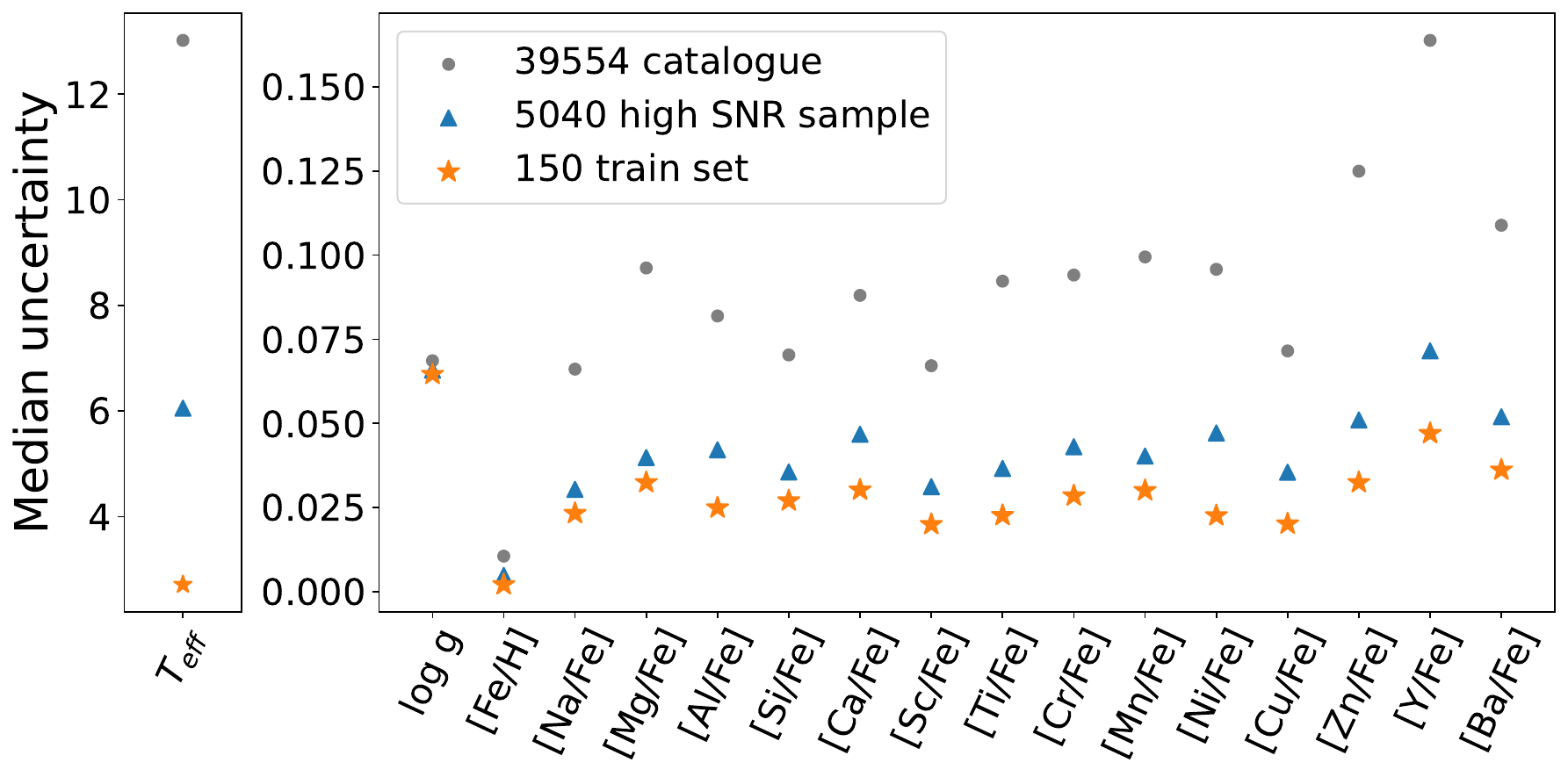}
    \caption{Median internal uncertainties per label as reported in GALAH DR3. In gray dots we show all 39\,554 solar twins from our selected catalogue. In blue triangles we show the \highSNRsample\ high SNR sample of solar twins. The orange stars represent the selection of 150 stars used as final training set (see discussions in Sect.~\ref{sect:final_trainset}). Uncertainties are smaller for the high SNR samples which are used for training. }
    \label{fig:median_uncertainties_per_label}
\end{figure*}

Among these spectra, some stars have repeated observations. We use that sample to assess the uncertainties in our results. Information about repeat observations can be found in Appendix~\ref{sec:repeat}. 

\subsection{Determining abundances with The Cannon} \label{sec:method}

We use the method \textit{The Cannon} to determine the new abundances. This is a data-driven approach that allows us to derive stellar labels (in our case, stellar parameters and chemical abundances) from stellar spectra. {In short, the code connects the flux of a spectrum at different wavelengths with a set of labels, by constructing a polynomial model with linear, quadratic, and cross-term coefficients for the labels.}  It was introduced by \cite{ness-15} and has since then been widely applied to stellar spectra \citep{galah-dr2, Casey-17, nandakumar-22}. One of its most practical strengths is that it does not use a physical but an empirical model of the spectra. This allows the method to obtain labels at comparable high precision compared with a standard derivation of abundances from e.g. synthetic spectral fitting, and it does it fast and computationally cheap.

{\it The Cannon} requires the existence of a subset of reference objects (in other words, a training set) which has well-determined stellar labels and must cover the parameter space sufficiently well and evenly. In a dataset like GALAH, it was found that there are too few metal-poor stars with well-determined labels to characterise well enough the metal-poor population observed by this survey \citep{galah-dr2}.  In our case, by focusing on the solar twins only, we have higher confidence to build a valuable training set for {\it The Cannon}. 

\begin{figure}

    \includegraphics[scale=0.17]{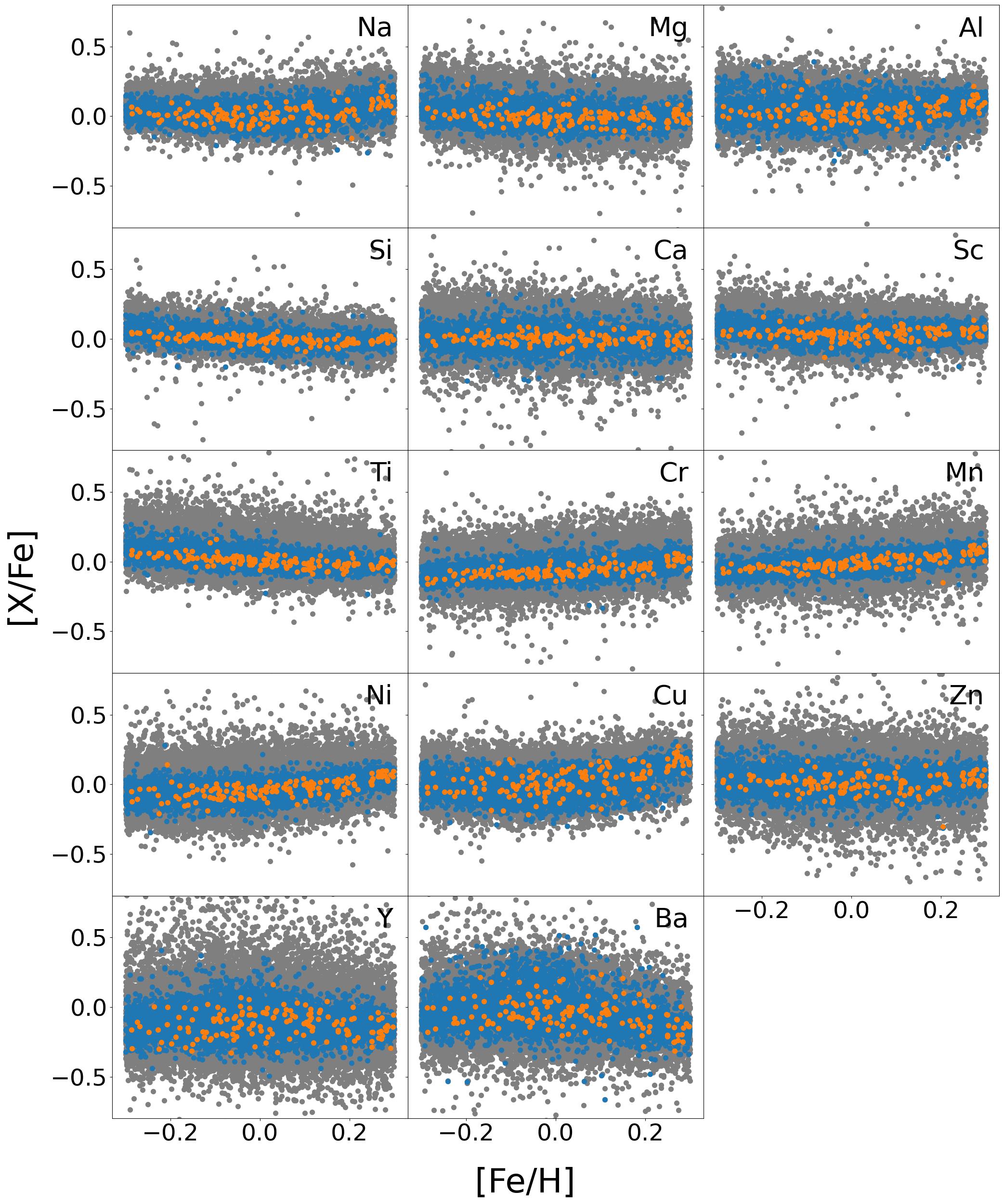}
    \caption{Panels of abundances in the format of [X/Fe] vs [Fe/H]. The corresponding abundance X is displayed in each panel.  In gray, all 39\,554 solar twins data. In blue, all \highSNRsample\ high SNR solar twins data. In orange, 150 high SNR spectra of solar twins training set selection (see Section~\ref{sect:final_trainset}). The entire dataset covers a wider range in abundances, but that could be due to higher uncertainties. Both train set and high SNR set have similar range in abundances.  }
    \label{fig:xfe_vs_feh}
\end{figure}

\subsubsection{Training set}

We are interested in training our model with a set of high quality spectra with accurate label  measurements. That model has to be used to generate new labels and uncertainties for the 39\,554 solar twin stars.  To select the best quality data we base our criterion on the SNR of the spectra, because GALAH~DR3 provides its most accurate and precise parameters and abundances for high SNR spectra \citep{GALAH-DR3}. We thus consider a set of spectra with SNR above 50 across all CCDs because  these are the highest quality spectra obtained in GALAH DR3 and thus provide the most precise results. Since we want to make new predictions on the labels of the whole solar twin dataset, we consider the full dataset of 39\,554 spectra as a test set.  There are 5\,144 high SNR spectra (hereafter called high SNR sample) for training and 39\,554 for test/prediction. These are different spectra of different stars. Among these 5\,144 high SNR spectra there are still some stars which have problems of normalisation or data reduction. We explored by eye all these stars and removed the bad spectra, reducing our training set to \highSNRsample\ stars.%

Fig.~\ref{fig:median_uncertainties_per_label} shows the median uncertainties in all labels for different selections of our sample. In gray we show the median internal uncertainties of the entire solar twin catalogue. In blue we plot the median uncertainties for all stars in the high SNR sample, and in orange we show the median uncertainties for a selection of a training set of 150 stars (Sect.~\ref{sect:final_trainset}). We can see how our training set has labels that are more precise than the rest of the catalogue for all labels except for surface gravity, which was predominantly estimated from non-spectroscopic features. Therefore, its precision is not dominated by SNR. 

{In Fig.~\ref{fig:xfe_vs_feh} we show the individual abundances as a function of the metallicity, defined as \feh, for the 39\,554 solar twins as well as the \highSNRsample\ high SNR solar twins subset and the 150 high SNR solar twin selected as training set (see Sect.~\ref{sect:final_trainset}) following the same colour scheme as Fig.~\ref{fig:median_uncertainties_per_label}.
We see the high SNR solar twin subset does not fully cover the parameter space of the entire catalogue.
The low SNR spectra might induce a spread in the abundances that is driven by the uncertainties. However, we observe the selected training of 150 stars set has a good coverage of the metallicity and abundances when compared with the high SNR sample.}

\subsubsection{Masks}
\cite{galah-dr2}  discussed that \Cannon\ had a better performance while using masks in the spectra for each label than when using the entire spectrum without filtering specific wavelength regions.  This means our \Cannon\ models are not performed in every pixel for every label but only on pixels that were selected to contain information about the label. That information is known from synthetic spectra \citep{galah-dr2}. Later on, these masks have been considered by \textsc{sme} to perform the fitting of observations with synthetic spectra to provide the abundances of all GALAH stars in GALAH~DR3. We used the same masks to make models for the different label predictions, although we comment that the masks are constantly being revisited in GALAH, therefore they might not be identical to those used by us here.

\section{The Cannon catalogue}\label{sec:catalogue}
In this section we present our results about the performance of our new catalogue of high precision abundances of solar twins in GALAH. Our catalogue of abundances can be downloaded from Vizier (link provided upon acceptance of paper). 

\begin{figure}
    \centering
    \includegraphics[scale=0.15]{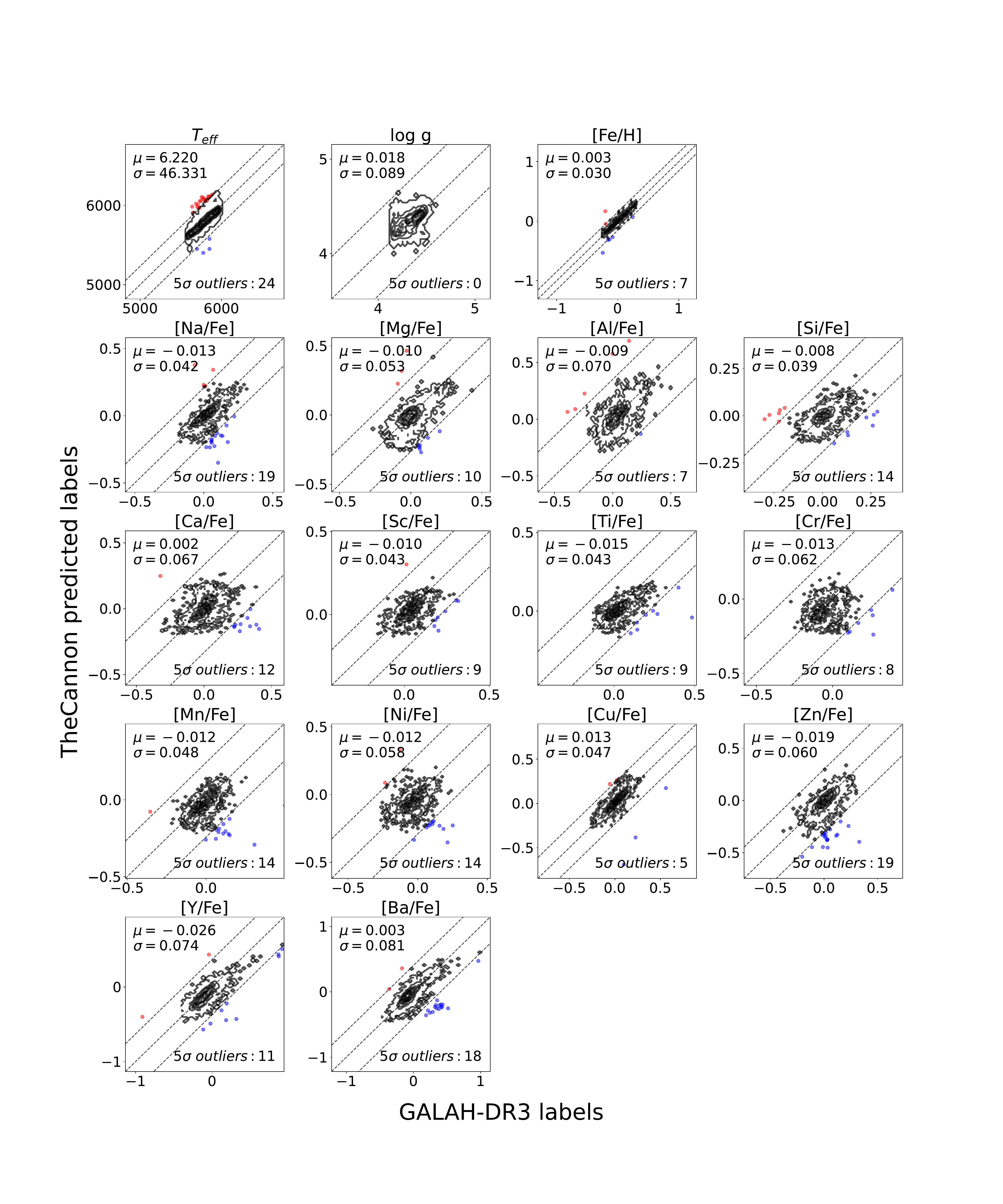}
    \caption{Comparison of labels obtained with \Cannon\ and GALAH-DR3 for stars with  $\mathrm{min(SNR)} > 50$. Each panel correspond to a different label. Outer dashed lines correspond to $5\sigma$ boundaries. Overestimates are plotted in red, underestimates in blue. The mean $\mu$ and standard deviation $\sigma$ of the difference between results are specified in each panel, as well as the number of outliers found outside each boundary.}
    \label{fig:1to1_highsnr_minsnr117_150spectra}
\end{figure}

\begin{figure}
    \centering
    \includegraphics[scale=0.15]{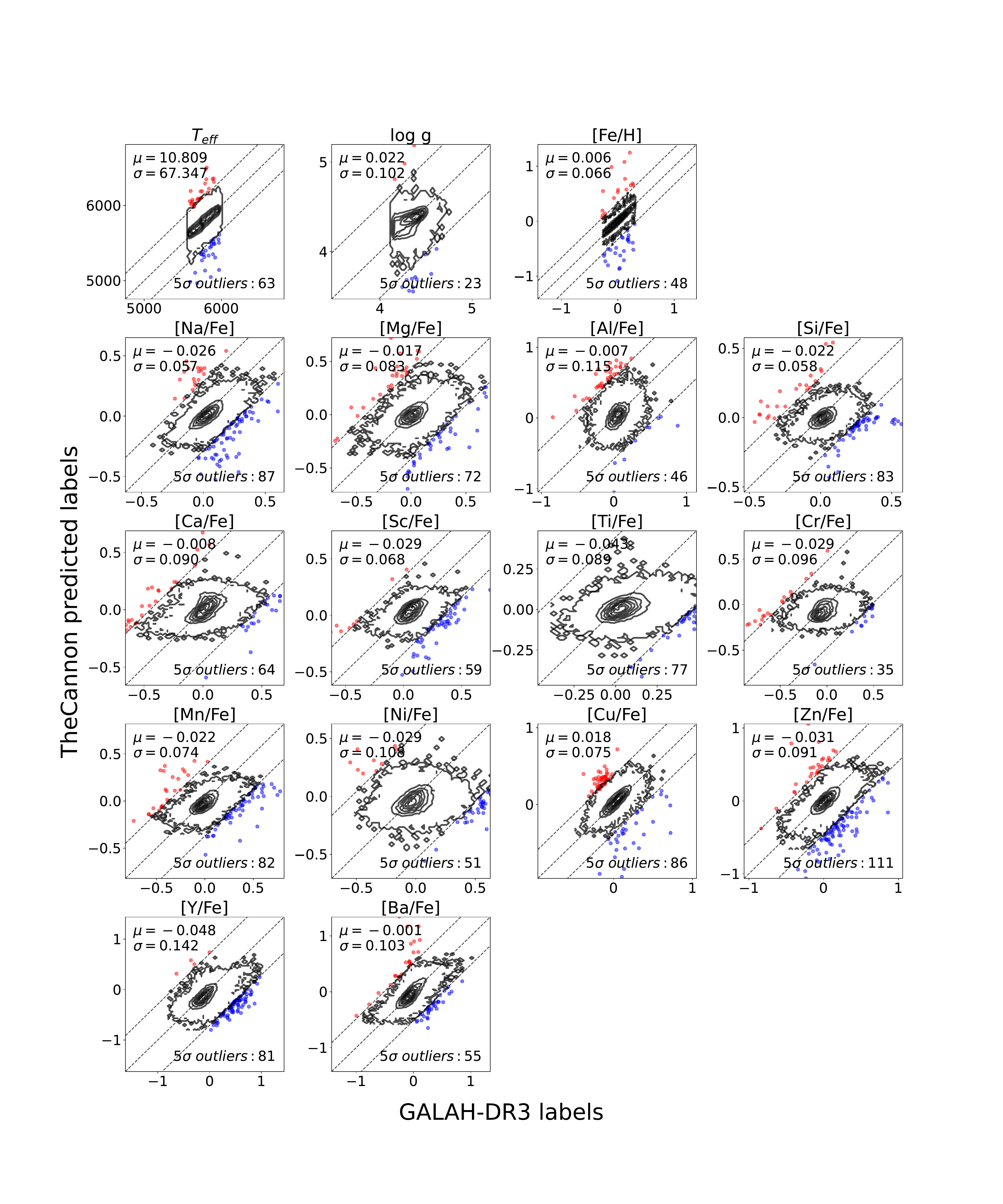}
    \caption{Same as Fig.~\ref{fig:1to1_highsnr_minsnr117_150spectra} but for stars with spectra of $\mathrm{min(SNR)} < 50$.}
    \label{fig:1to1_lowsnr_minsnr117_150spectra}
\end{figure}

\subsection{Comparison with GALAH DR3}

We use the \Cannon\ Model 3, namely the model built with 150 solar twins of minimum SNR of 117 across all CCDs (See Appendix~\ref{sec:results}).  In order to validate our results we look at the agreement of the predicted values over the 39\,554 solar twin catalogue at high and low SNR, as well as the uncertainties reported by our \Cannon\ model and compare them to the current GALAH~DR3 reported uncertainties. 

In Fig.~\ref{fig:1to1_highsnr_minsnr117_150spectra} we show the comparisons of our labels and the GALAH DR3 results for stars in the high SNR sample. For stellar parameters \teff\ and \feh\ we have a good agreement within 50~K and 0.03~dex, respectively,  with some exceptions of overestimates in the $5\sigma$ boundary for \teff. However for \logg\ we are not able to consistently recover the labels, obtaining a scatter in the one-to-one relation of 0.1~dex which is very large considering the small range of surface gravities in our sample. We recall that GALAH~DR3 does not derive surface gravities from the spectra because these spectra are not sensitive to surface gravity. It is thus not surprising that the agreement is poor.  For the chemical abundances we have a good agreement overall. However, we can notice some outliers for Al and Cr in the sense that we overestimate the abundances for a few metal-poor stars and underestimate the abundances for some other metal-richer stars.

In Fig.~\ref{fig:1to1_lowsnr_minsnr117_150spectra} we display the same results but for stars with low SNR (below 50). As expected, the number of outliers increases,  with stars predicted to have effective temperatures outside the solar twin range. It is expected that the comparison will be worse in this case, as we know the measurements obtained by the pipeline of GALAH~DR3 are more uncertain for low SNR spectra. 

We still obtain a good agreement for Na, Mg, Cu, Zn, Y, Ba, although with a higher dispersion than the high SNR stars. However for Ca, Cr we observe two trends in our prediction.  The model makes near-flat predictions for metal-poor and metal-rich stars, resulting in two groups in the comparisons, with the model underestimating the abundances for these  groups. For Al we observe a higher slope in the one-to-one relation, e.g. we overestimate this abundance. For Ti we obtain a flat prediction, namely all the stars in the low SNR regime have solar-like Ti abundance.  In our final catalogue we remove all the outliers outside $5\sigma$ boundaries found in both high and low SNR. {This translates into further removing 89 high SNR stars and 749 low SNR stars, obtaining a final catalogue of 38\,716 solar twins.} 

\begin{figure*}
    \centering
\hspace{-3cm}
    \includegraphics[scale=0.3]{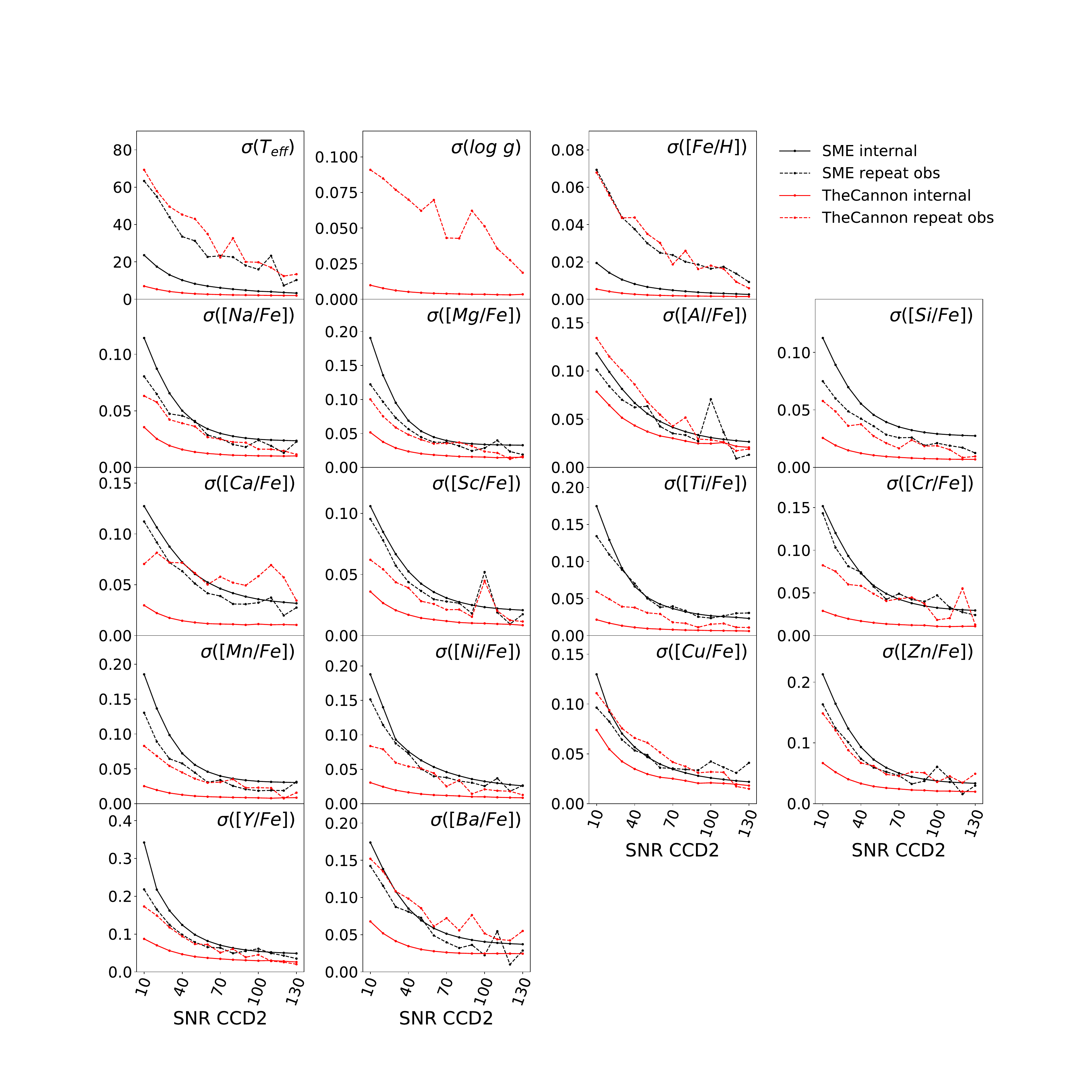}
    \hspace{-1cm}
    \caption{Standard deviation of uncertainty labels as function of SNR CCD2. In black, \textsc{sme} uncertainties. In red, \Cannon\ uncertainties. Solid lines represent the internal uncertainties given by the covariance matrix of the fitting by both \textsc{sme} and \Cannon. Dashed lines represent the uncertainties of repeat observations.} 
    \label{fig:sigmas_minsnr117_150spectra}
\end{figure*}

\subsection{Comparison of uncertainties}

We use the internal uncertainties obtained by our model and compare them to the ones already given by GALAH~DR3. We also use the repeat observations sample (see Appendix.~\ref{sec:repeat}) where we made predictions with the model and compare them with the uncertainties of repeat observations given by GALAH~DR3. 
In Fig~\ref{fig:sigmas_minsnr117_150spectra} we summarise our findings, we have both internal uncertainties (in solid lines) and repeat observations uncertainties (in dashed lines) for GALAH~DR3/\textsc{sme} in black and our \Cannon\  in red as function of SNR. We removed the \textsc{sme} uncertainties for \logg\ since the estimation for this label does not rely on spectral fitting like our \Cannon\ model does. 

In general the internal uncertainties reported by our model are below the ones reported by GALAH~DR3.  For GALAH~DR3 they tend to increase as SNR decreases. Our model also predicts labels more uncertain at lower SNR, but the difference in uncertainties between high and low SNR is smaller than for GALAH~DR3.  For the uncertainties obtained from repeat observations however, our results are comparable to GALAH~DR3 for all SNR ranges.  The uncertainties of repeat observations for the stellar parameters \teff\ and \feh\ are similar for both pipelines as function of the SNR, and are generally higher than the internal uncertainties. For the chemical abundances we find some cases, where our model has higher uncertainties than SME, especially  at high SNR.  A notable example is Ca, where our \Cannon\ model reports higher repeat observations uncertainties at high SNR.  We note that in GALAH~DR3 a more detailed approach in masking telluric lines was made \citep{GALAH-DR3} but here we used a less refined mask which likely contains more telluric features. 

For Cu and Zn differences are negligible, which could be due to the fact that in both procedures the masking of the spectra is the same therefore the methods consider the same information from the spectra. The opposite difference is found for Si, Sc, Ti, Cr, Ni, Y where in general our uncertainties of repeat observations are lower than GALAH~DR3.  In particular, or Si and Mn our masks have more pixels than \textsc{sme} in GALAH DR3 which included an additional filtering due to blending but were not labeled as such in the masks. Our \Cannon\  model does not seem to be affected in terms of precision suggesting that perhaps the further filtering in the GALAH~DR3 masks was too strict for solar twins, removing pixels. 
For Ti we also observe that \Cannon\ obtains more precise abundances. \textsc{sme} determines Ti separately from neutral and ionised Ti lines,  whereas \Cannon\ takes all lines together.  It is  thus expected to have better precision in our model because more pixels are used.

In general, at lower SNR \Cannon\ performs better since it is always using information from the whole spectrum, whereas \textsc{sme} applies further filtering in masking the detected lines in each spectrum \citep{GALAH-DR3}. This effect may be reflected in the higher internal uncertainties reported at low SNR. When there is more information in lines available in the spectra \Cannon\ will use this information. If we are using the same amount of pixels as SME, the repeat observations uncertainties are comparable since they are fitting essentially the same features. 

\begin{figure}
    \centering
    \includegraphics[scale=0.18]{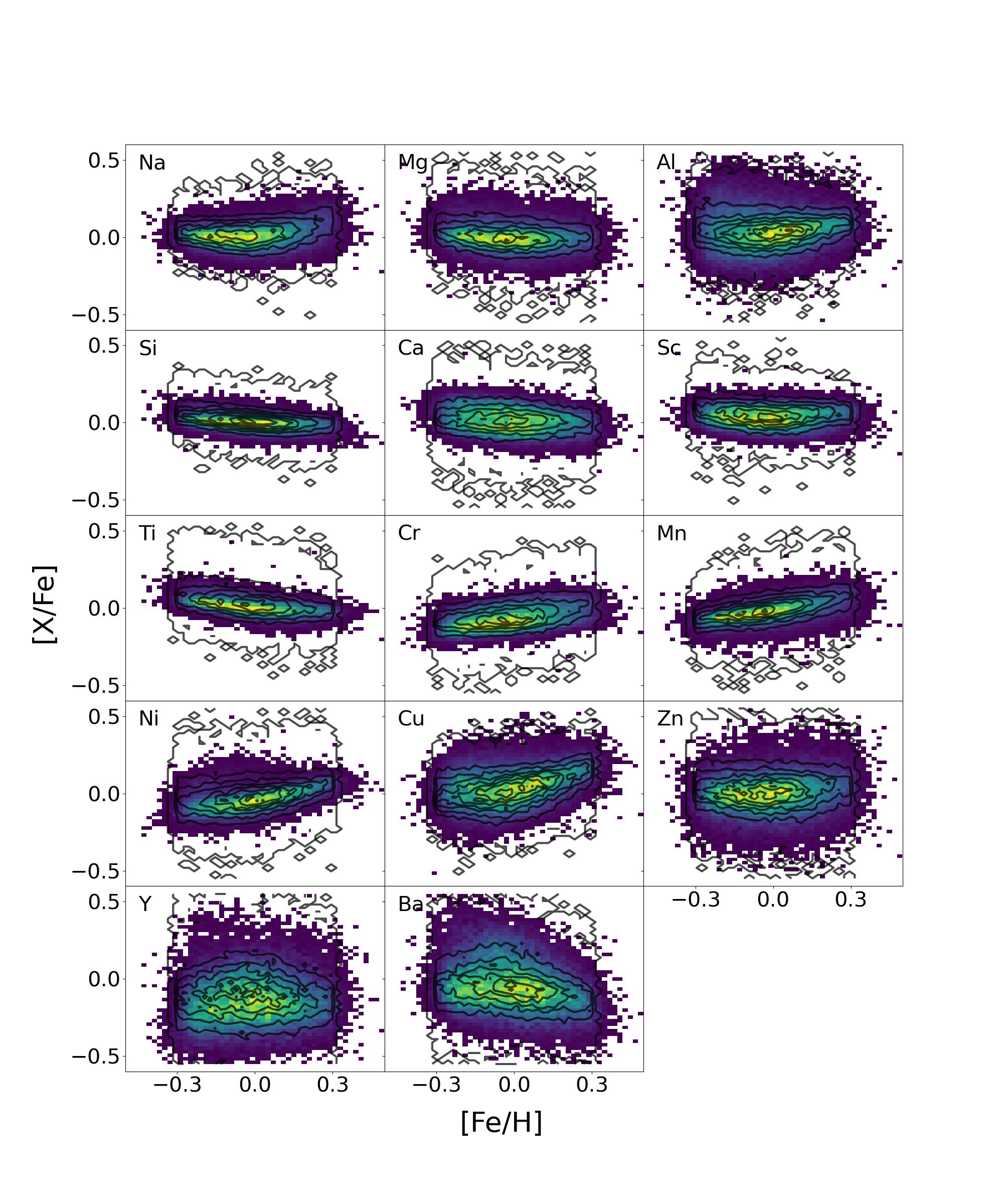}
    \caption{Individual abundances as a function of metallicity for the solar twins analysed in this work. Coloured density plots correspond to our labels as determined using \Cannon\ while contours delineate the distribution of GALAH DR3.}    \label{fig:xfe_feh_minsnr117_150spectra}
\end{figure}

\begin{figure}
    \hspace{-0.5cm}
    \includegraphics[scale=0.18]{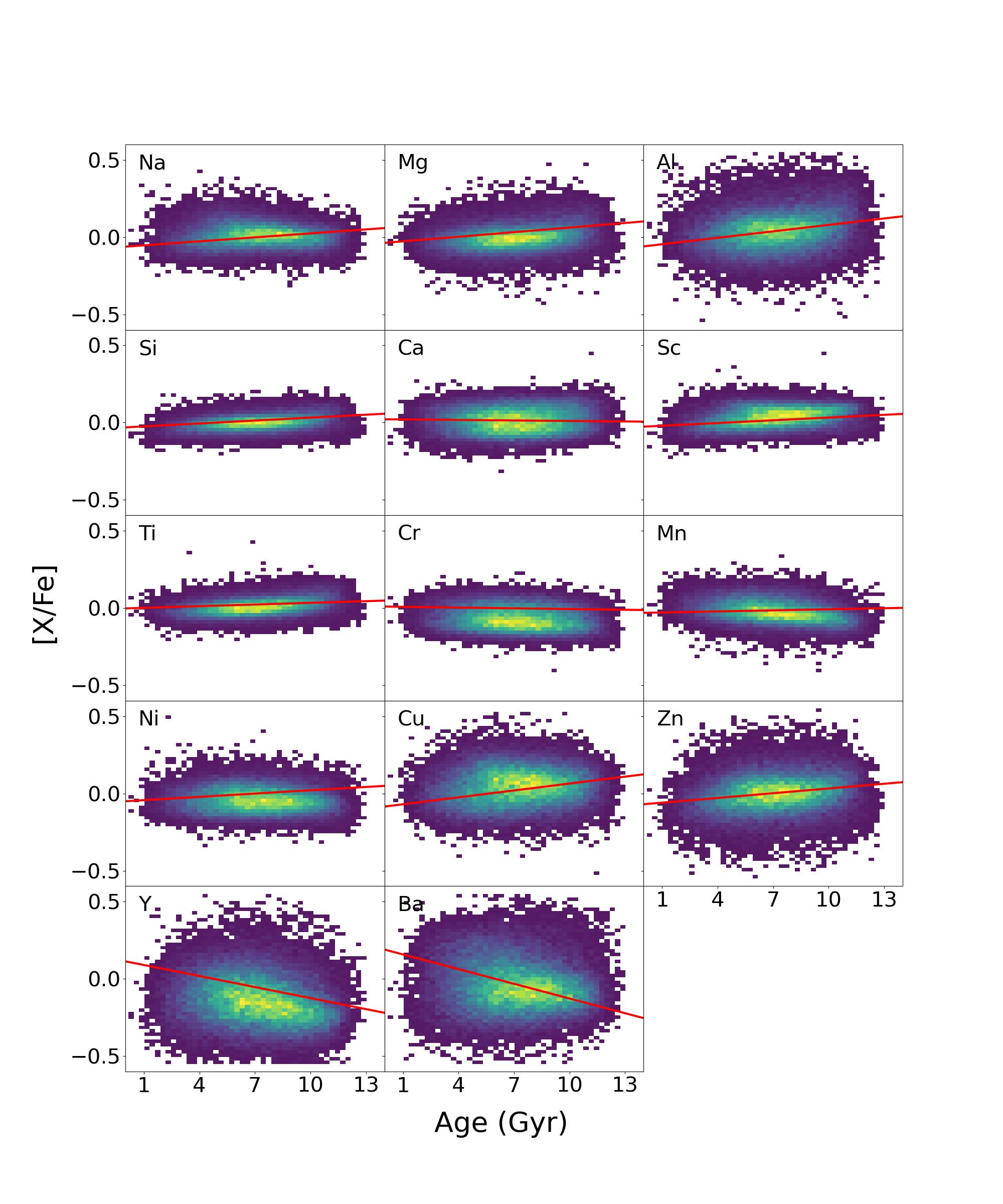}
    \caption{Individual abundances as function of stellar age for the solar twins analysed in this work. Coloured density plots correspond to our labels as determined using \Cannon. Red solid lines correspond to the linear regression fits of abundance-age trends found by \citet{Spina-16} and \citet{Bedell-18}. Age estimates are taken from GALAH~DR3 \citep{GALAH-DR3}.} 
    \label{fig:xfe_age_minsnr117_150spectra}
\end{figure}

\subsection{Abundance distributions}

Fig.~\ref{fig:xfe_feh_minsnr117_150spectra} shows the individual abundances as a function of metallicity for \Cannon\ values coloured as density plots and for GALAH~DR3 as contours. We can see how the precision in terms of the dispersion improves significantly,  with exceptions for Al, Cu,  Zn, Y and Ba.  
Al, Cu and Zn have very weak lines which makes them a very difficult element to measure even for \Cannon. The internal uncertainties are comparable for these elements as discussed in the previous section.   Y and Ba are elements that are expected to present a large scatter. Stars in binary systems that had an AGB companion might have been polluted by s-process elements such as Y and Ba that were produced by the AGB star \citep{escorza-19}. Based on Gaia DR3, however, we do not find a clear signature of binarity for stars with higher [Ba/Fe] and [Y/Fe] abundance ratios in their RUWE\footnote{RUWE stands for Renormalised Unit Weight Error and is determined in Gaia to account for uncertainties in the astrometric solution. It is considered that a value above 1.4 could indicate that the uncertainty in the solution is higher due to the non-single nature of the star, which would change its position in the sky. } parameter  or the uncertainties in the radial velocities. If such high Y and Ba stars were  in binary systems, their separations would be large and their periods long.  

Elements such as Mg, Si, Ca and Ti are formed predominantly in SNII progenitors \citep[see][and references therein]{kobayashi-20}. In this sample,  we do not find stars that might have formed from an $\alpha-$enhanced gas such as the thick disk, deducing that this sample is composed predominantly by thin disk stars.  Iron-peak elements such as Sc, Cr, Mn and Ni follow the trends observed in other higher resolution and higher precision studies, such as \cite{Adibekyan-12}, \cite{bensby-14}, and \cite{Battistini-15}.  

Fig.~\ref{fig:xfe_age_minsnr117_150spectra} illustrates our values as a function of age.  Here the age corresponds to the values reported by GALAH~DR3. 
We note these ages are not fully consistent with our new parameters since they are estimated with the GALAH-DR3 parameters. A new derivation of ages is beyond the scope of this paper. Here we aim to provide an illustration of the type of studies that could be performed with the entire usage of our catalogue. 

As above, the coloured density plots represent our values determined with \Cannon. The red line corresponds to the linear regression fits of the abundance-age trend determined by \cite{Spina-16} and \cite{Bedell-18}  who performed a high precision spectroscopic analysis of solar twins using high resolution. We qualitatively obtain consistent trends for all elements. There are minor offsets, such as Cr, but that was already discussed in \cite{GALAH-DR3}. Since we train with GALAH~DR3 values, it is expected that the offset remains here. 

{With respect to Fig.~\ref{fig:xfe_age_minsnr117_150spectra}, we also want to stress that the underlying training set of GALAH~DR3 is subject to significant selection effects and systematic parameter inaccuracies. This includes the overestimation (or systematic clumping) of stellar ages of stars around $2~\mathrm{Gyr}$, due to the missing separation of young star isochrones in our solar twin parameter space. We also expect to sample more intermediate age thin disk stars from the underlying set of stars due to their relative abundance within the GALAH selection function (neglecting the Galactic plane and sampling within magnitude ranges).}

\section{Phylogenetic trees with GALAH solar twins}
\label{sec:applications}

In this section, we use our new catalogue in the construction of phylogenetic trees. 
For our purpose, we select from the catalogue two groups of solar twins with different orbit eccentricity. More specifically, we select the 100 stars with lowest eccentricities and 101 stars with highest eccentricities in the sample. These values come as valued-added information in GALAH and are derived from Gaia DR3 data (see Sect.~\ref{sec:data} for details).  
Our goal is to test how the phylogenetic trees built using our measurements tell us about their relatedness, and how our measurements help in this goal compared to the GALAH-DR3 ones.  For this experiment, we compare the phylogenies constructed from both datasets.  The stars selected for the analysis are listed in Tab.~\ref{tab:selected_stars}. In that table, we are including the stars ID as labelled in the tips of our trees, in addition to the Gaia DR3 IDs for further references, and their ages and eccentricities as downloaded from the valued added catalogues of GALAH-DR3.

A natural question arises here about our choice of stars. Indeed, there is no particular reason to choose our sample above any other sample. But we need to stress that the time complexity of the NJ algorithm is $O(n^3)$ \citep{yang2014molecular}, where $n$ is the number of tips, so applying NJ algorithm on 40,000 stars might be computationally impractical. This implies we need to make choices of smaller sets of data, and focus our scientific aim to a particular question. In this paper, we aim to compare the phylogenetic signal between our \Cannon\ and the standard GALAH-DR3 abundances.  

\subsection{Building trees with the Neighbor-Joining algorithm}

{Trees are built following the methods used in \cite{Jackson-21} and \cite{Jofre-17Phylo} by using the classical method called Neighbour-Joining (NJ), a computationally fast agglomerative cluster algorithm proposed by \cite{Saitou1987} that iteratively joins nodes (in our case, stars) closely related by a given pairwise distance matrix. This matrix of distances has size $N \times N$ with $N$ being the number of stars considered in the analysis, and the distance is calculated as a {\it Manhattan Distance} between the chemical abundances of each star in the pair. The tree reconstruction is made in a greedy way by iteratively joining a pair of nodes from the distance matrix that minimises the {\it Q criterion}. For a tree with $r$ nodes, the pair $\left(i,j\right)$ is joint by minimizing 
\begin{equation}
    Q_{ij} = (r-2)d_{ij}- \sum_{k=1}^r(d_{ik}+d_{jk}), \mathrm{for}~ i<j\leq r, 
\end{equation} 
\noindent where $d$ is the distance between the pairs \citep[See Sect. 3.3.3 of ][for more details of this terminology]{yang2014molecular}. Then  a new node is created and is called an internal node, which acts as predecessor of the two joint nodes.  The procedure continues by recomputing the distances of the remaining nodes to the new internal node, removing the rows and columns in the distance matrix related to the two joined nodes and adding the ones related to the new one. This reduces the dimension of the distance matrix by 1. The process iterates again until the distance matrix is of size $2 \times 2$, joining the two remaining nodes and returning the built tree.}

We build our distance matrix using our selected sample of stars from both eccentricity groups (see Tab.~\ref{tab:selected_stars}). To compute the distance, we consider a vector of measured abundances for each star, and use the {\it Manhattan Distance} which is the absolute difference of two vectors. This means that we are using chemical distances for the stars in our sample in a similar way to \cite{Jofre-17Phylo}. In our case, we perform a further selection of the abundances, namely those whose age-abundance trend is monotonic in Fig.~\ref{fig:xfe_age_minsnr117_150spectra}. This is known to increase the additivity of the distance matrix, hence making NJ trees that are closer to the true phylogenetic tree \citep[][see also Eldridge et al in prep for additivity in stellar abundance data]{retzlaff2018phylogenetics}. We thus exclude Ca, Cr and Ni since their age-abundance trends are flat, thus not evolving in time. Since these abundance ratios do not change in our sample, they add noise in our tree reconstruction \citep[]{yang2014molecular, Jackson-21}.

To account for the uncertainties in the data, we build NJ trees from distance matrices computed by empirically sampling a random value out of a normal distribution for each of the abundances, centered at the reported measurement and with a standard deviation of its reported uncertainty.  We build 2\,000 trees by sampling the abundances according to their uncertainties and study their distribution. 

From these trees, we select the best tree to be the one which has highest node support. To do so, we follow the process used in biology that searches the {\it maximum clade credibility} (MCC) tree out of a sample of trees. Clades correspond to a group of nodes that includes all the descendants of a common predecessor node in the tree. 
We note that with trees being built empirically we cannot immediately assume that the trees make evolutionary sense, hence an internal node can not be immeditately associated with a  clade. Using simulated data is needed to learn the prospects and limitations of interpreting clades and evolutionary histories from empirical trees \citep{debrito23}.  In the MCC, each clade (or node with all the descendant nodes in our trees) in each sampled tree is given a score that reflects the fraction of times that the same pattern appears in all the sampled trees.  If the clade occurs for all the trees then the support value is 1 (100\%).  This indicates high consistency in the data for that topological relationship. The product of these scores is defined as the tree score, so that the MCC tree is the one with the highest tree score.
Here we employ this method to select our best tree and evaluate its robustness, despite not being able to ensure that our nodes and branching pattern can be directly related to clades (see more discussions below). 

\begin{figure}
    \centering
    \includegraphics[scale=0.38]{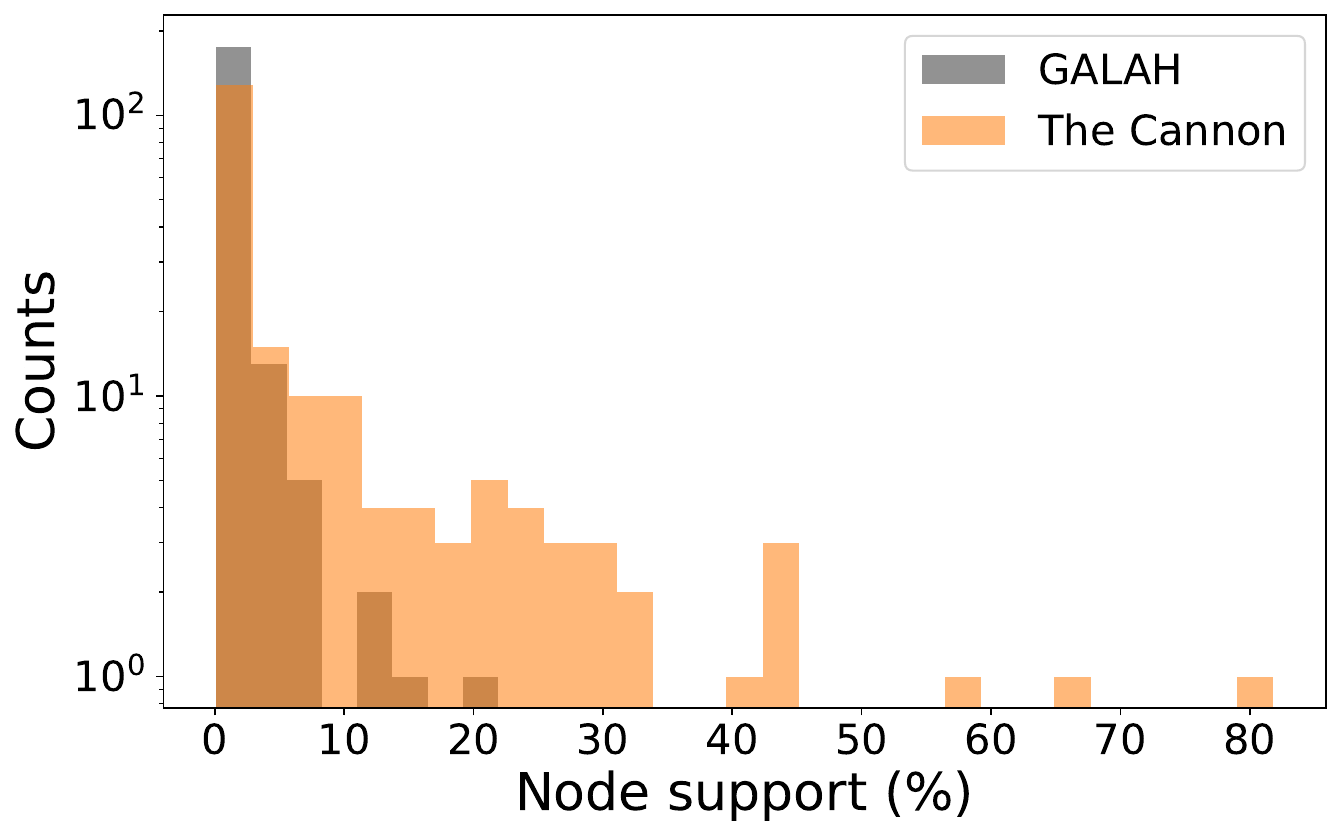}
    \caption{Node support percentage for MCC trees of Fig.~\ref{fig:mcctrees_ecc}. In gray, support percentages for the MCC tree built using GALAH~DR3 data. In orange, support percentages for the MCC tree built using \Cannon\ data.}
    \label{fig:mcctree_cladesupports}
\end{figure}

The distribution of support values for the nodes is shown in Fig.~\ref{fig:mcctree_cladesupports}. In gray and orange we have the support values for the MCC tree built using GALAH and \Cannon, respectively. By comparing the distributions we observe that \Cannon\ MCC tree is overall better supported. The GALAH  highest support value in the GALAH MCC tree is 20\%, meaning that every clade seen in the MCC tree only appears in at most 20\% of the remaining sampled trees. However, even though the \Cannon\ MCC tree has more support, the overall values do not exceed 50\%. This means that even at the high precision in the abundances of our new catalogue, the trees are overall poorly supported.

The root of the tree is the basal split that separates the most distant (in an evolutionary way) object from all the rest.  The NJ algorithm produces unrooted trees because in this tree reconstruction method there is no evolutionary model considered and hence no way predict the ancestral state in the relationships of our stars.  Hence, even though we are able to apply the MCC method to find the most supported tree, we are not able to attribute a clade in our trees as a group of nodes that includes all the descendants of a predecessor node in an evolutionary context. In our case, it is more appropriate to refer to possible groups in the trees as clans instead of clades.

\begin{figure*}
    \hspace{-1cm}
    \includegraphics[width=0.49\textwidth]{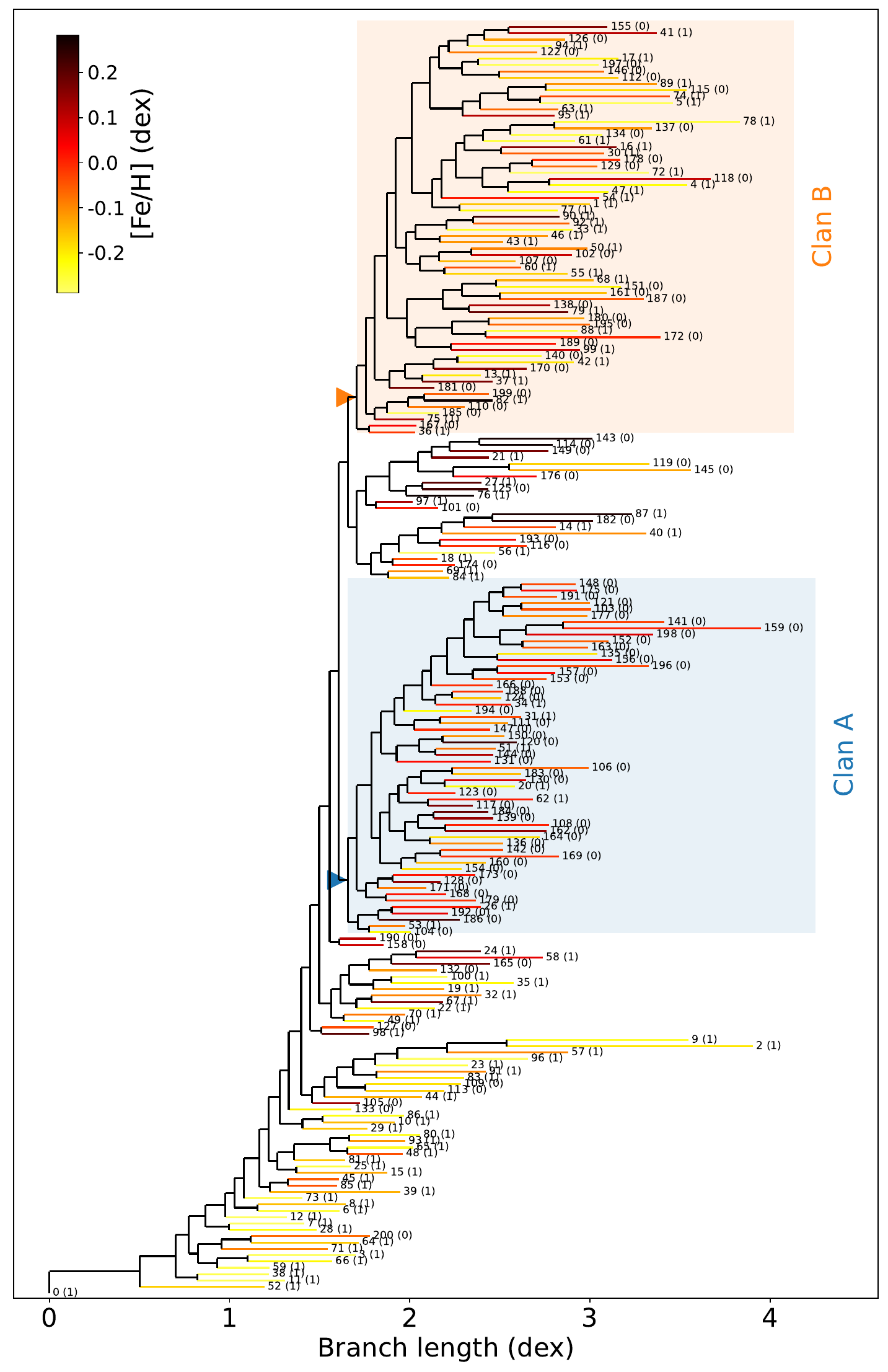}
    \includegraphics[width=0.49\textwidth]{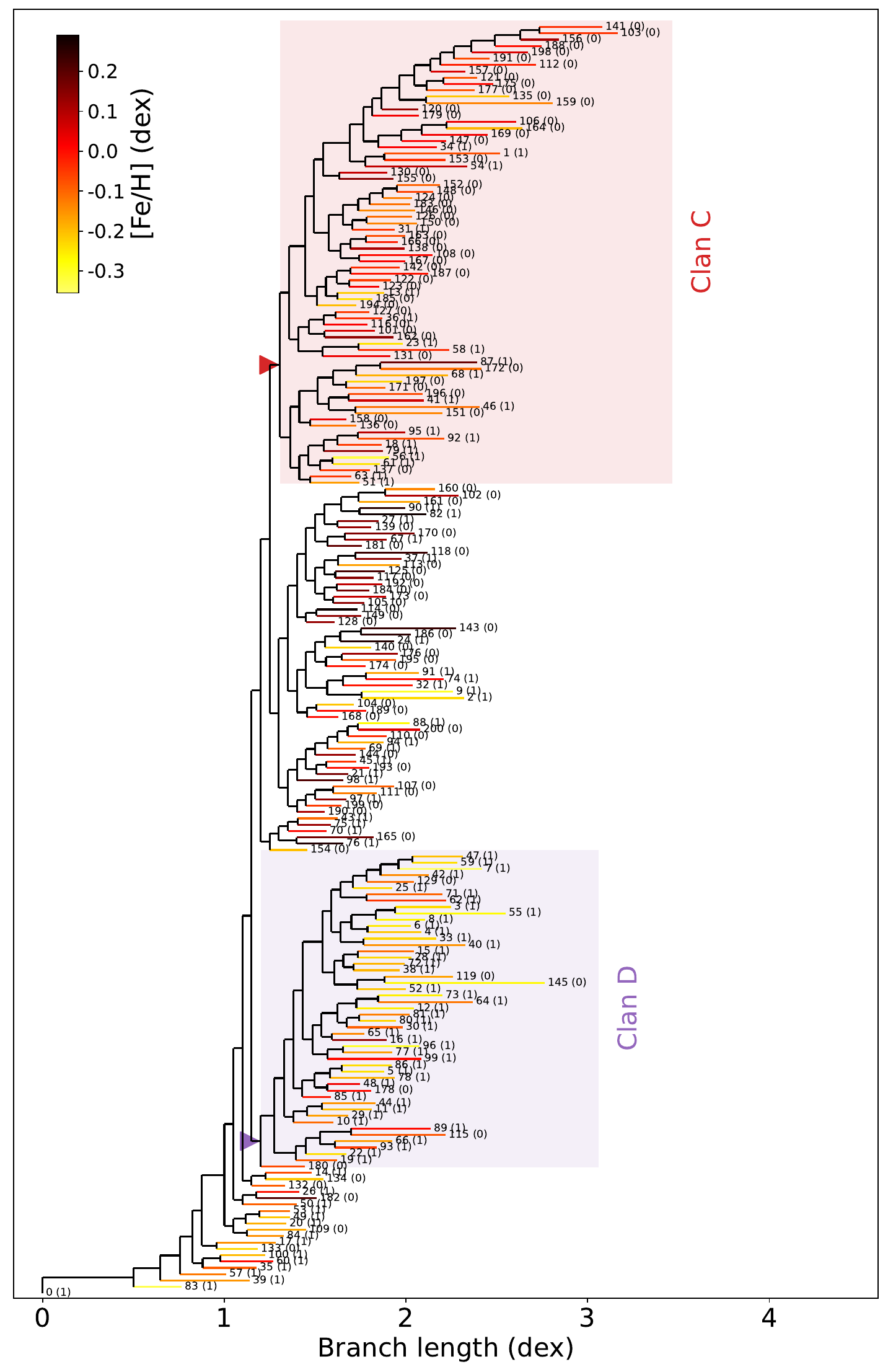}
    \caption{Maximum Clade Credibility (MCC) trees from a sample of 2\,000 trees for 201 stars of different eccentricity groups with tips coloured by eccentricity. On the left, MCC tree obtained with GALAH labels. On the right, MCC tree obtained with \Cannon\ labels. Stars in selected eccentricity groups are enumerated in the tips with values between 0 and 200, 0 being the fixed branch placed at the most eccentric solar twin in the catalogue  (eccentricity of 0.63), 1-100 being the following most eccentric stars and 101-200 referring to the less eccentric stars in the catalogue (see Tab.~\ref{tab:selected_stars}). Selected clans A-B and C-D can be seen from the coloured areas of GALAH and \Cannon, respectively.}   
    \label{fig:mcctrees_ecc}
\end{figure*}

\subsection{GALAH vs Cannon trees}

In Fig.~\ref{fig:mcctrees_ecc} we show the MCC trees built from our sample stars. For better visualisation of our trees, we choose the tip corresponding to the star labelled with ID 0 as our reference star. This means, the tree is displayed in a way that all branch lengths are visualised with respect to ID 0. That star is the one with highest eccentricity in our sample. Its GaiaDR3 ID is 5396076243592498944, and it has an eccentricity of 0.63. This allows us to study the relationship of all stars with respect to that high eccentric one.  

We stress that the tree is not rooted. We could have chosen any star as our reference star for visualizing the relationships, e.g. the oldest one, the most metal-poor one, or simply a random one. The tree is displayed with respect to that star for better comparison between trees, and the branching order that follows the node with that reference star does not necessarily mean an ancestral-descendant relationship because we do not have a root nor any prior or model about the evolutionary history of our selected stars. In this case we chose the most eccentric star of our sample simply because its possibilities to come from a region beyond the Solar neighborhood is higher than a star with circular orbit, implying its chances to be more evolutionarily distant to the rest are higher. Furthermore, the dynamical properties of the stars depend on Gaia data only, and are independent of the spectroscopic parameters (as well as ages and chemical abundances). This allows us to use the same star as reference for both datasets we consider in building the tree (our \Cannon\ abundances and GALAH-DR3).

The left panel of Fig.~\ref{fig:mcctrees_ecc} shows the MCC tree obtained using GALAH abundances and the right panel shows the MCC tree obtained using \Cannon\ abundances. Both trees have the branches coloured by the metallicity as obtained from the corresponding catalogue. The parenthesis in each tip of the trees  correspond to the eccentricity group, where 0 represents circular orbits and 1 represent more eccentric orbits. Specific information about eccentricities and ages of our stars can be found in Tab.~\ref{tab:selected_stars}.

By comparing the topologies of these trees, we see some similarities. In both cases, the eccentric stars are located close to the star ID 0, and after a few splits we are able to observe two main branches. Within these branches, we select clans for further studies, which we label Clan A and B for the GALAH tree, and Clan C and D for the \Cannon\ tree. We will discuss these clans with more detail later on. 

{We also observe that the length of the tip branches differ between the trees. In fact, the tip branch lengths of the GALAH tree are larger than the ones with \Cannon. Some GALAH branches reach 4 dex while Cannon ones reach 3 dex. In fact, the GALAH tree has tip branch lengths which are larger than the inner branches. This is an indication of a {\it hard} tree \citep{yang2014molecular}, which are trees prone to errors. In the GALAH tree, the difference between two stars, reflected by the sum of horizontal branch lengths connecting the two tips along the tree, is dominated by the tip branch lengths rather than the internal branch lengths, dominating over the hierarchical structure in the tree. This means that more of the chemical differences among the stars is being explained by their tips than any one of the internal branches.   This is also noticed in \Cannon\ tree but to a lesser extent. 

In any case, we observe that both trees chemically cluster together stars from both eccentricity groups, but with the \Cannon\ data the grouping is more resolved. In the GALAH tree, Clan A (highlighted with blue in Fig.~\ref{fig:mcctrees_ecc}) contains mostly the stars in circular orbits which are primarily metal-rich, while Clan B (in orange in the figure) contains a mix of stars. When using our catalogue, we see that Clan C (enclosed in red in the right hand panel of Fig.~\ref{fig:mcctrees_ecc}) contains mostly low eccentric metal-rich stars.  Clan D has (in violet in the figure) stars with high eccentricities and rather metal-poor.  

It is worth to comment on the selection of our clans. Between Clan C and Cland D there is a branch of stars that is more similar to Clan C than Clan D but the internal branches are short compared to the length of the tip branches,  and the topology is overall more balanced. That branching pattern is indeed similar similar to a random tree, lacking phylogenetic signal \citep{debrito23}. 



\begin{figure}
    \hspace{-1cm}
    \includegraphics[scale=0.35]{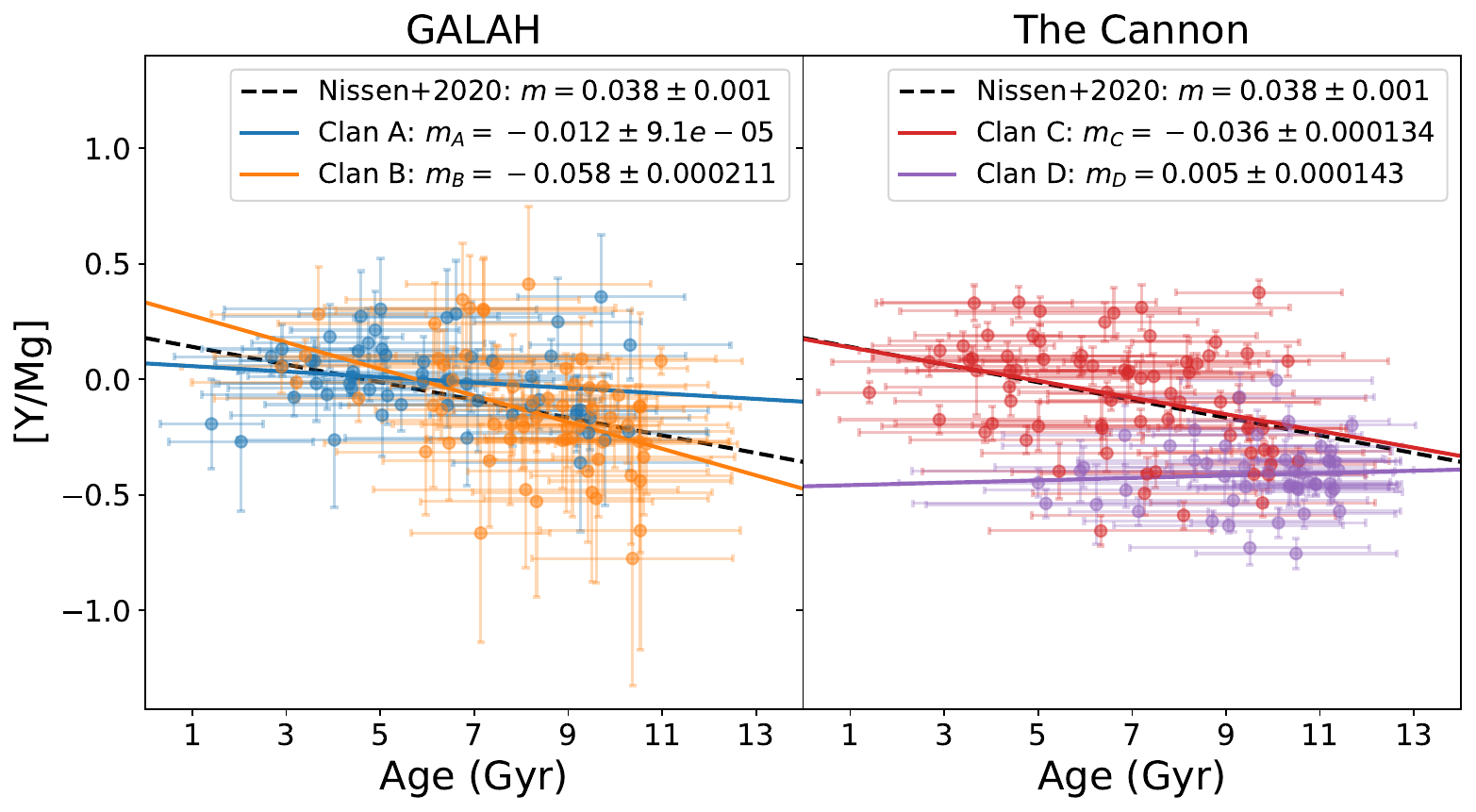}
    \caption{Age-[Y/Mg] relations of the stars in the clans selected from the trees of Fig.~\ref{fig:mcctrees_ecc}. Linear regression fits are drawn to quantify the slopes in these relations. Trend found by \citet{nissen-2020} displayed as a dashed black line. [Y/Mg] errors taken from the internal uncertainties reported by GALAH~DR3 and our \Cannon\ catalogue. Age errors taken from GALAH~DR3 \citet{GALAH-DR3}.} 
    \label{fig:age_ymg_clades}
\end{figure}

\subsection{Astrophysical interpretation of the selected clans}

{We now look into the Clan A and B from the GALAH~DR3 tree and Clan C and D from the \Cannon\ tree.  To do so, we explore the trend of age and [Y/Mg], commonly referred to as {\it chemical-clock}. Indeed, in the analysis performed on solar twins by \cite{Nissen2015} a tight relationship between age and [Y/Mg] was found. This trend was explained with the argument that yttrium, which is an element produced by AGB stars, increases with increasing Fe, while Mg, which is an element produced by SNII, decreases with increasing Fe. Since Fe increases with time, this difference in dependency with Fe causes a strong dependency of [Y/Mg] with age. 

After that study, several works have studied the applicability of this trend considering different kinds of stars, finding that solar-metallicity giants in the solar neighborhood behave similarly to the solar twins \citep{Slumstrup17, casamiquela21} but at lower metallicities, this relationship might weaken \citep[][Vitali et al in prep]{delgado-mena, Casali-2020}. It is also suspected that this relation is subject to systematics in the age determination \citep{Berger22}.  Further studies have found this relation might not hold for stars outside the solar neighborhood \citep{casamiquela21}, which can be explained by the fact that this relation has a strong dependency of the the star formation rate, which is different at different birth radii \citep{ratcliffe23}.

Considering that  $\mathrm{[Y/Mg]} =  \mathrm{[Y/Fe]} - \mathrm{[Mg/Fe]}$, the errors in the abundance ratio used in this work are computed as the quadratic sum of the errors reported for both abundances. Age estimates as well as age errors are taken from \citet{GALAH-DR3}. In Fig.~\ref{fig:age_ymg_clades} we show the age$-\mathrm{[Y/Mg]}$ trends for the selected clans in Fig.~\ref{fig:mcctrees_ecc} following the same colours.  We compute a linear regression fit of the stars in each group, and plot the fit with a line of the same colour as the corresponding clan. The legend indicates the value of the slope and the uncertainty of the fit. The dashed black line corresponds to the slope of the linear fit found by \citet{nissen-2020}, for reference.

{We see that the age$- \mathrm{[Y/Mg]}$ trends have different slopes for the different clans. However, among all the trends, it is remarkable the agreement of the slope in the fit found trend found for Clan C and the \cite{nissen-2020} fit. Clan C is the group which is composed by a majority of low eccentric stars using \Cannon\ abundances. 
We note that the stars analysed by \cite{nissen-2020} have eccentricites normally below 0.1 \citep[see also][]{Nissen2015, Jofre-17Phylo, Jackson-21}.  It is encouraging to find that the stars in Clan C follow so well the chemical clock found in other studies on solar twins.  

Interpreting this finding in terms of the phylogenetic nature of this group is however tricky. As shown recently by \cite{ratcliffe23}, the age$- \mathrm{[Y/Mg]}$ relationship found by \citet[][and references therein]{nissen-2020} can be interpreted only by considering that birth radii also plays a fundamental role. It is only possible to explain that a sample of stars with a restricted metallicity range in the solar neighborhood can have a range in ages if they come from different galactic radii. Like this,  each star traces a different star formation rate and reaches the same [Fe/H] at different timescales. The fact that Clan C is composed of stars that are mostly on circular orbits, but have a range in ages, suggests that the oldest stars might have migrated from inner regions. We would expect that old stars with circular orbits that have not migrated should be significantly more metal-poor. These are not here because we have only selected solar-metallicity stars for our study.   

Our selection in metallicity is however not too restrictive. In fact, we have a range of 0.6 dex in metallicity in the sample, and Clan C contains stars of all metallicities,  indicating that some ISM evolution at the solar radius must be present. Disentangling which stars are product of the inheritance of the ISM at the solar radius and which have migrated or are visiting due to dynamical heating (higher eccentricities) is tricky since all these processes are mixed in the disk \citep{Feltzin20} and evolve as time passes \citep{Aumer16, Bird2021, Lu2022} 

The slope of Clan D is rather flat, but that could be due to the fact that the stars in Clan D are predominantly eccentric. These stars might have originated from different Galactic radii and are less exposed to have a shared history. Most of them are also old, making the resulting fit biased. It is currently believed that stars originating from different galactic radii might have different age$-\mathrm{[Y/Mg]}$ relations \citep{Casali-2020, casamiquela21}. Furthermore, the trend and its relationship with birth radius evolves with time. For oldest stars, this ratio could have been flat across the Galaxy \citep{ratcliffe23}, and this is consistent with our findings. 

\begin{figure}
    \hspace{-0.5cm}
    \includegraphics[scale=0.35]{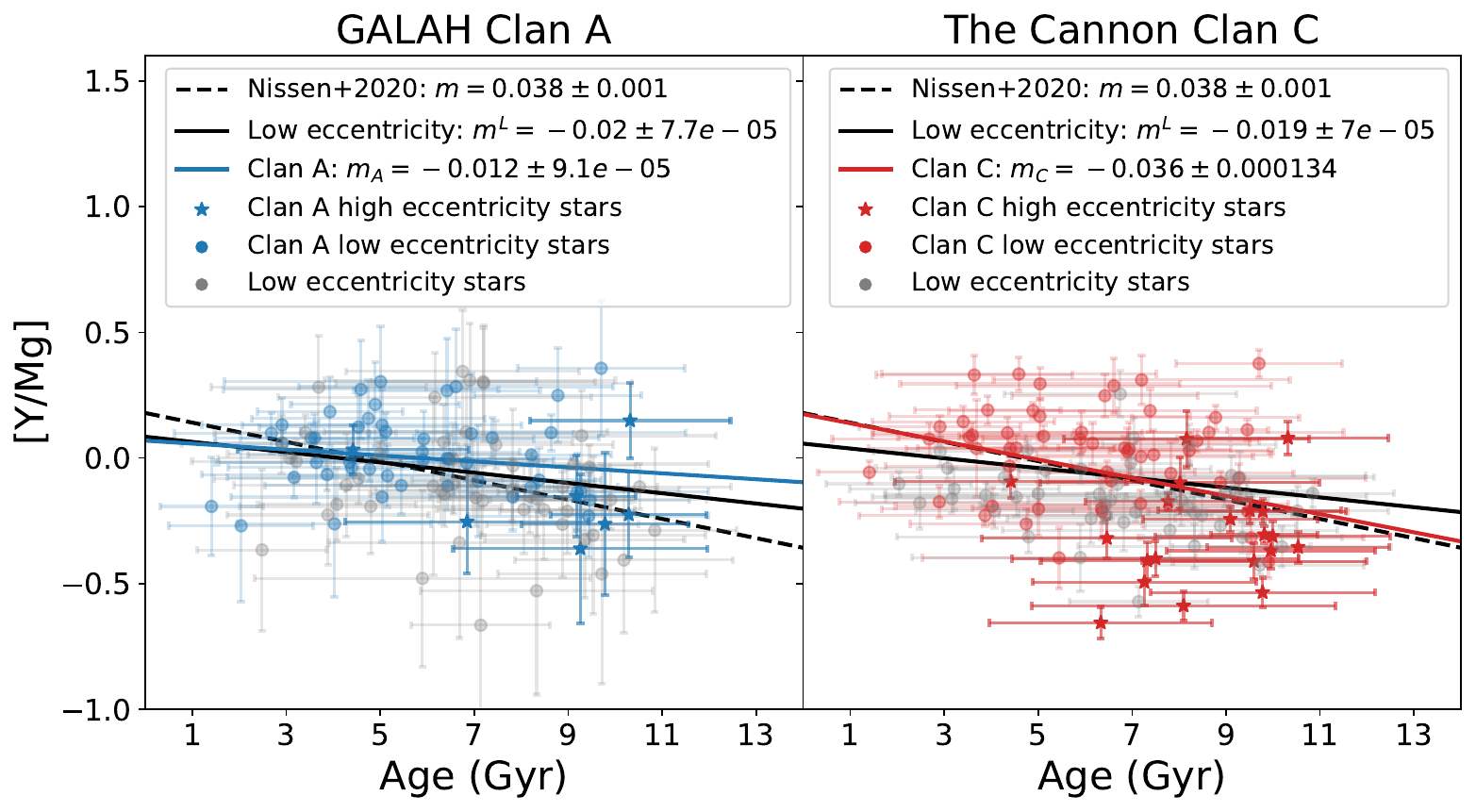}
    \caption{Age-[Y/Mg] relations of the stars in Clans A (GALAH~DR3) and C (\Cannon) selected from the trees of Fig.~\ref{fig:mcctrees_ecc}. Linear regression fits are drawn to quantify the slopes in these relations. A linear fit using all low eccentricity stars selected for the analysis is displayed as a solid black line. Trend found by \citet{nissen-2020} displayed as a dashed black line. [Y/Mg] errors taken from the internal uncertainties reported by GALAH~DR3 and our \Cannon\ catalogue. Age errors taken from GALAH~DR3 \citet{GALAH-DR3}.} 
    \label{fig:age_ymg_clades_compare}
\end{figure}

For the GALAH Clans A and B we see a  smaller difference when considering the uncertainties. 
Moreover, Clan A has an age-[Y/Mg] trend which is flatter than Clan B, which is the opposite to what we find with the \Cannon\ abundances. It is thus  hard to explain that Clan A, which contains predominantly low eccentric stars, deviates more from the \cite{nissen-2020} relation than Clan B, which contains a mix of stars. This might be a consequence of higher uncertainties in the abundances of GALAH~DR3. 

{A natural question may arise as to whether we are able or not to recover the same trends by simply doing  dynamical cuts for the sample stars. To answer this question we consider Clans A and C, because they are the groups that mainly contain low eccentric stars. We compare the slopes of their chemical clocks with the ones we would obtain if we considered all the 100 low eccentricity stars in the sample. 
Fig.~\ref{fig:age_ymg_clades_compare} summarises this result.  On the left panel we plot again in blue the stars belonging to Clan A and the blue solid line is the resulting linear fit to these stars. Since Clan A does not only contain low eccentric stars, for better illustration of our findings we plot with different symbols the stars with circular orbits (in circles) and eccentric orbits (in stars). The resting stars in low eccentricity orbits are plotted with gray circles, and the linear fit is shown in the black solid line. Finally, for reference, the chemical clock fit found by \cite{nissen-2020} is shown with the dashed line, as before.  All slopes are specified in the legend. For Clan C, which considers the abundances obtained by us with \Cannon, the same information is shown in the right-hand panel, and the stars are coloured in red.

As we see in Fig.~\ref{fig:age_ymg_clades_compare}, by just considering the low eccentricity stars we are not able to find trends that are consistent within the literature, even considering the uncertainties. This might be because we are missing important older stars which now are on less circular orbits. A cut on dynamical properties only is therefore not sufficient since the selected groups are incomplete \citep{soubiran05, Hawkins15}.  When using the tools provided by phylogenetic analyses, we can chemically identify different groups of stars and find patterns that could be associated to their shared history. We stress that we obtain this result only for our new high precision abundances, demonstrating also the importance of having very high precision abundances for a better selection of stellar populations. }



\subsection{Discussion}

The NJ  algorithm is essentially a clustering algorithm, not particularly different to others available in the literature \citep[see e.g.][for descriptions and discussions of different clustering algorithms used in chemical data]{Ratcliffe20}. What makes the NJ algorithm however attractive here is that first we do not have to specify the number of clusters we aim to find, unlike other fast clustering algorithms such as K-means. This is important when studying the relationships and shared history of a group of stars as a whole, where we are not primarily interested in finding groups but in studying the way in which the entire system is ordered and how this order might tell us something about their evolutionary history. Second, the NJ algorithm is not designed to only cluster the data, but to visualise the amount of divergence between pairs of objects. This translates into branch lengths that have a meaning of difference. NJ trees therefore are not expected to have the objects aligned at the tips of the tree, in contrast to other dendograms obtained by clustering algorithms such as DBSCAN or HDBSCAN \citep{casamiquela21tagging} or other agglomerative methods \citep{Ratcliffe20}. The procedure to define the branching pattern only depends on the distance matrix. Other clustering algorithms require to specify parameters of closeness and density in the parameter space, which cannot be constrained in an objective way \citep{casamiquela21tagging}. In fact, here we are not primarily interested in finding clusters and quantifying the number of clusters and their properties, but to visualise how the data is structured in their hierarchical order. Because of the heritable information we consider to build the trees, that order can be used to interpret shared histories, which is the essence of phylogenetics.  

We comment on the poor support of our trees, where in the best case we still have a large majority of nodes with a support below 50\% (see Fig.~\ref{fig:mcctree_cladesupports}). Considering that the NJ algorithm takes a distance matrix as input and joins the elements that are closest to each other, by construction it will generate a tree that will reflect the hierarchical order of the distance matrix. But when the range of differences is small, a perturbation of that distribution given the uncertainties will imply a very different tree. The distance method for tree reconstruction becomes uncertain if the distances are too small for the entire sample \citep{yang2014molecular}.  

Using solar twins for this kind of study might also add further challenges in the interpretation of the results. Having a sample of a restricted range in metallicity  makes it impossible to trace back the population that was formed from the pristine gas (e.g. with no metals). The fact that all stars are metal enriched tells us that we are studying a population from a stage in which considerable chemical enrichment already happened.  Stars of different ages and different [Y/Mg] but similar [Fe/H] might well be formed at different galactic environments and arrive to the solar neighborhood through a dynamical process that can be radial migration or heating.

Given the different timescales in the pollution of Mg, Y and Fe into the ISM, it is not straightforward to interpret a clan which has a tight relationship of [Y/Mg] and age but has a limited range in [Fe/H]. Is it that the NJ is simply clustering stars of different birth radii which trace different evolutionary histories of the ISM? Perhaps the AGB stars, which live longer than the progenitors of SNII,  had enough time to radially migrate and pollute the ISM with Y at a different location than their birth place in contrast to the pollution of Mg by their massive siblings \citep{Johnson}. It is possible that the ISM has a shared evolutionary history at a wider range in radii due to migration. The interplay between migration, heating, blurring and mixing in the ISM of the Milky Way is still poorly understood \citep{Feltzin20}. Phylogenetic methods might thus offer an interesting opportunity to learn more about these processes. 

We also need to remind ourselves of the selection effect in our sample, where the selection of solar twins systematically underrepresents young stars.  As explained in \citet{Sharma2018},  the age estimation for both the youngest and oldest stars via the isochrone fitting is biased towards intermediate ages.

There is another source of uncertainty here, which is the fact that by considering loose stars in the disk, we can not rule out the possibility that two stars will have the same origin (e.g. be siblings of the same star formation episode of the same molecular cloud). The NJ will place these two stars in two different tips of leaves in the tree (by construction), but they in fact represent one leave. Without a previous selection of stars belonging to distinct star formation episodes, the NJ algorithm will fail in ranking the leaves, because any hierarchical order obtained for stars from the same star formation episode with be driven by the errors in the abundances and the intrinsic dispersion of such populations. We also have to keep in mind that there will be a noise because of the ISM inhomogeneities that does not reflect evolution \citep{Kos21, Ness22}. The hope is that noise is less than the change due to evolution \citep{Manea22}. In that sense, nodes of poor support could also be used to identify co-natal stars.  To test these possibilities, simulated data is more suitable, because in that case we know with certainty the origin of the data \citep{debrito23}.

{As seen from Figs.~\ref{fig:mcctree_cladesupports}, \ref{fig:age_ymg_clades} and \ref{fig:age_ymg_clades_compare},} GALAH data still does not have the precision in the abundances required to perform a robust phylogenetic study. However, GALAH data has been central for this analysis, because \Cannon\ uses the best GALAH results to perform a re-analysis which then allows us to apply phylogenetic techniques in other GALAH stars. Considering that future data releases of GALAH are expected to increase in precision, the prospects of phylogenetic studies in GALAH are indeed very promising.

\section{Conclusions}\label{sec:conclusion}

In this paper we have performed a systematic application of the machine learning algorithm \Cannon\ to a set of solar twins observed and analysed in GALAH DR3 \citep{GALAH-DR3} with the aim to provide a catalogue of high precision abundances of {38\,716} solar twins and use a set of this catalogue for a phylogenetic study of GALAH data.  Other scientific applications of high precision abundances of solar twins include setting constraints on planet engulfment processes  \citep{Bedell-18, Tucci-Maia-19} or the level of homogeneity in star formation regions such as open clusters or wide binaries \citep{Liu-19, Hawkins-20, EspinozaRojas-21}. Therefore, our  catalogue can be used to explore these subjects, in addition to our primary intention to perform a phylogenetic analysis.  

In the systematic application of \Cannon\ for the generation of this catalogue, we investigated the impact of the labels considering different training sets. We first varied the size of the training set, and then studied the label recovery when removing outliers. This analysis helped us to conclude that a training set with 150 GALAH stars of SNR > 117 was sufficient for predicting precise labels of stellar parameters and 14 chemical abundances of solar twins observed with GALAH.

Our results agree within {50K in temperature, 0.09 dex in \logg, 0.03 in [Fe/H], and in 0.05 dex in abundances} with GALAH for stars SNR > 50. For lower SNR the results agree less, within 60 K in temperature, 0.1 dex in gravity, 0.07 dex in metallicity and 0.1 dex in the other abundances. This is expected considering that also GALAH data are more uncertain for lower SNR.  The internal uncertainties of our model at lower SNR do not significantly increase compared to high SNR results. The consistency of predicted labels for repeat observations remains comparable to the GALAH ones, provided both GALAH and \Cannon\ used the same pixels to extract the information. 

Our new catalogue allows us to perform phylogenetic studies on the solar neighborhood that require high precision abundances. We analysed 200 stars separated in two eccentricity groups, namely a group with circular orbits and another one with orbits of eccentricity around 0.4, and we compared the trees obtained for these groups using GALAH and \Cannon\ abundances. In both cases, we were able to find clans which were distinct in eccentricities, with one clan notably grouping the stars with circular orbits.  While the node support in the \Cannon\ tree is higher than for the GALAH tree, the overall support in both trees is not outstanding. This is expected for a sample of stars which are so similar to each other that the hierarchical differences between stars is very small and thus uncertain. It is also possible that many of these solar twins are tracers of the same star formation episode, causing a conflict in the tree shape which is forced by the neighbor joining algorithm. To truly study the support and amount of information carried out in chemical abundances, simulated data should be used instead, even if simulated data is prone to systematic uncertainties \citep{debrito23}. 

The trees still allowed us to study the astrophysical nature of the clans found. To this aim, we compared age-[Y/Mg] relation (e.g. chemical clock) of the clans with the literature \citep{nissen-2020}, obtaining a remarkable agreement for one of our clans with our \Cannon\ abundances, but no agreement with the clans found in the GALAH tree. The agreement between the chemical clock of \cite{nissen-2020} and ours was obtained noting that the clan included mostly stars with circular orbits, but some older stars with eccentric orbits as well. Indeed, the  age-[Y/Mg] relation found for only stars with circular orbits does not agree with the chemical clock obtained by \cite{nissen-2020}, because of the lack of old stars with circular orbits. A phylogenetic tree can thus help to identify the stellar family that traces the chemical evolution of the solar neighborhood, despite these stars having changed their orbital properties during their lifetimes. The abundances, however, need to be of very high precision.

Our work demonstrates the promising future of galactic phylogenetics, in which we can use large spectroscopic surveys like GALAH with machine learning to improve the chemical abundances which then can be used as input for phylogenetic analyses and so reconstruct the history of our home galaxy, the Milky Way.

\section*{Acknowledgements}
This work has been funded by Millennium Nucleus ERIS NCN2021\_017 and ANID Master's Degree Scholarship 22230932. We acknowledge FONDECYT Regular Grants 1200703 and 1231057. PJ thanks the Stromlo Distinguished Visitor Program at the ANU. E.J.J. acknowledges support from FONDECYT Iniciaci\'on en Investigaci\'on 2020 Project 11200263.  We warmly acknowledge Sara Vitali, Danielle de Brito Silva and Scarlet Elgueta for fruitful conversations throughout the development of this project. 

\section*{Data Availability}
Our catalogue of high precision solar twins from GALAH will be made public at the Virtual Observatory through Vizier. The spectra used in this work can be directly obtained from the GALAH survey. 

\bibliographystyle{mnras}
\bibliography{bibio.bib}   

\appendix{

\section{Repeat Observations}\label{sec:repeat}

\begin{figure}
    \centering
    \includegraphics[scale=0.15]{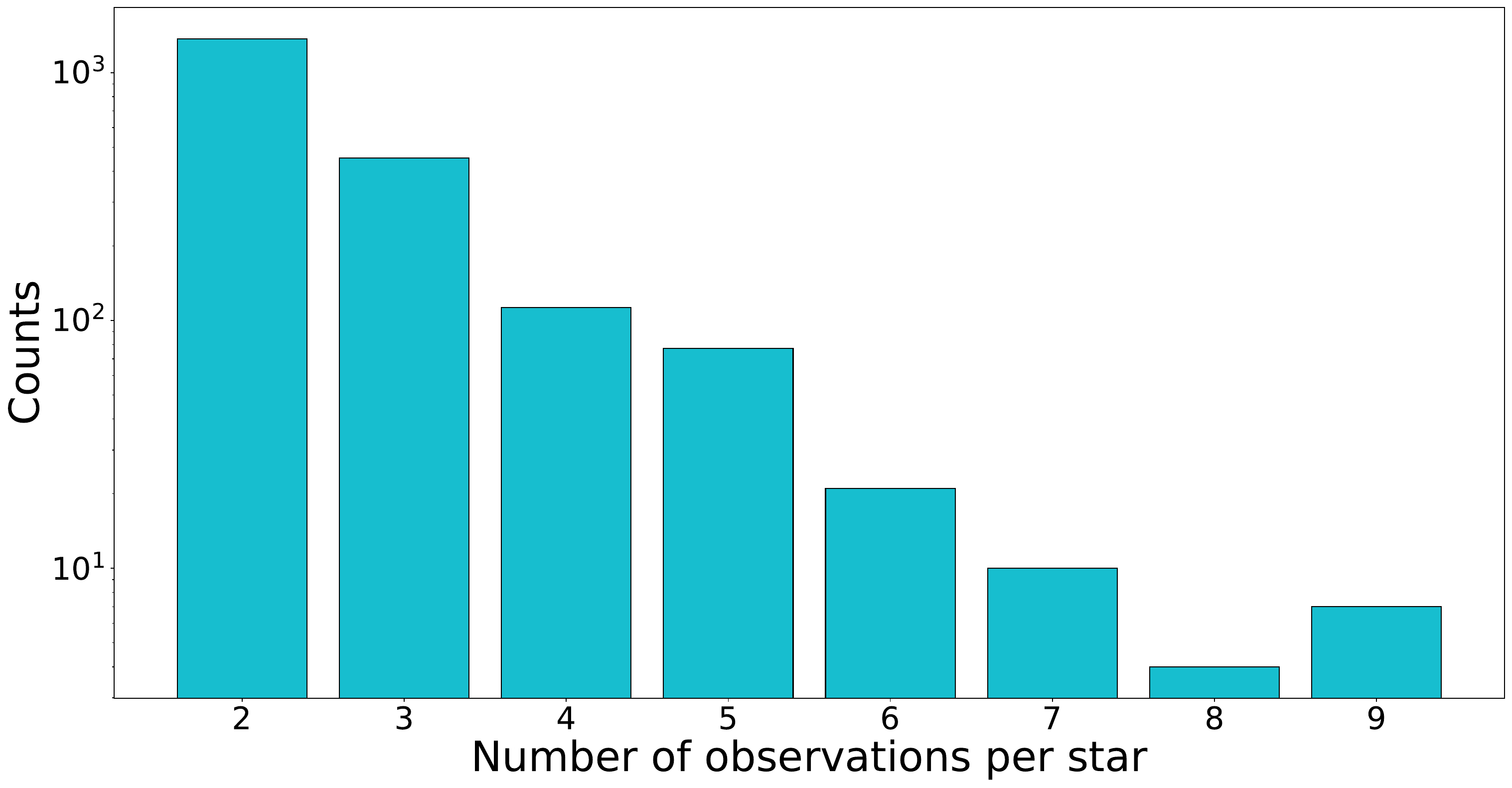}
    \includegraphics[scale=0.15]{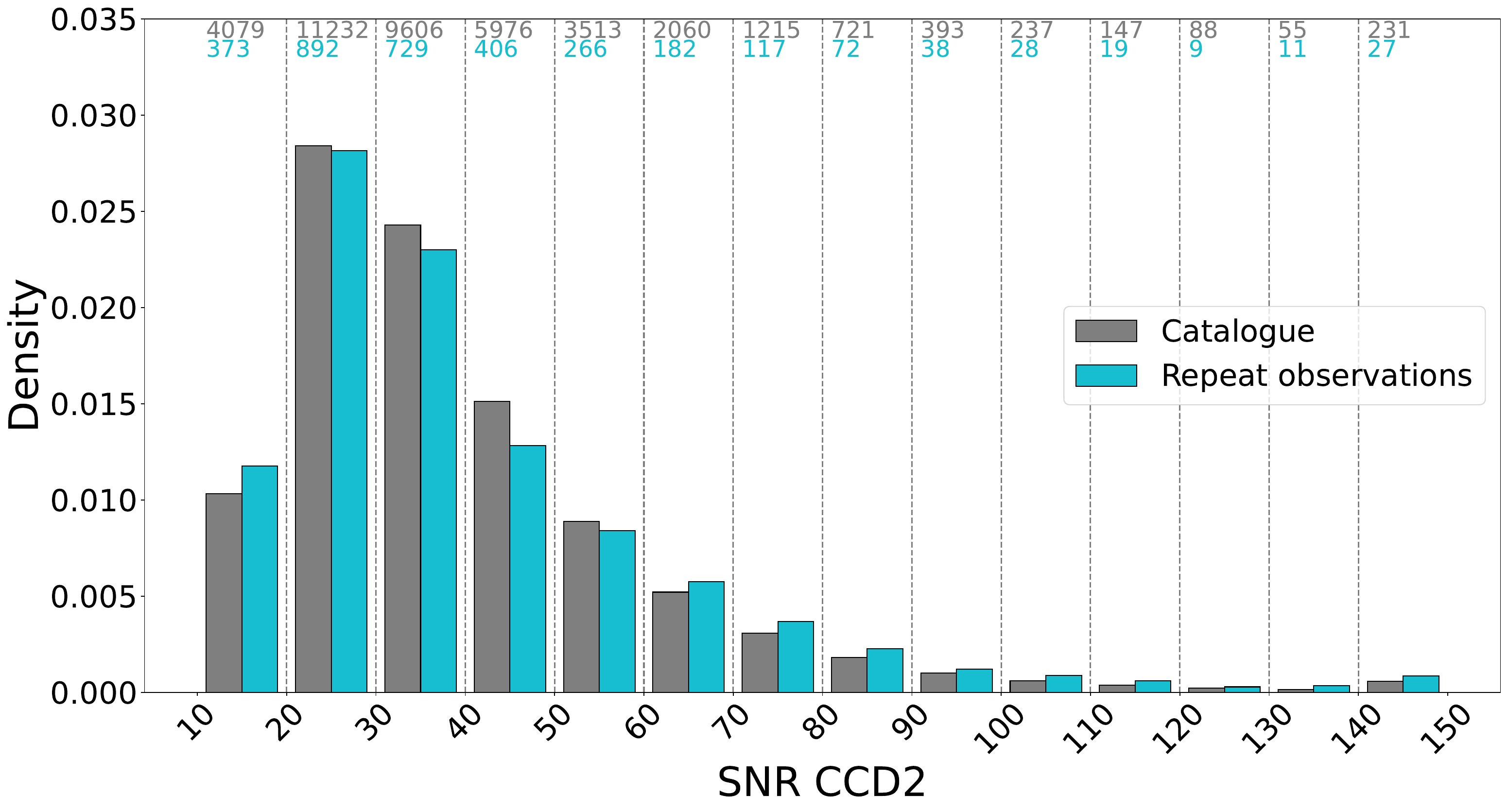}
    \caption{Top panel: Histogram of number of observations for repeated stars in the e entire GALAH DR3 sample. Bottom panel: Density histogram as function of SNR in CCD2. In gray, all 39\,554 solar twins catalogue. In cyan, repeat observations of solar twins. Raw counts above each bin on the top of the figure. Note for the last SNR bin we collapsed the all catalog data with SNR above 140.}
    \label{fig:histograms}
\end{figure}

To evaluate our results, we consider repeat observations for some stars. There are 12\,945 spectra of repeat observations in GALAH~DR3. 
Of those, we select the repetitions with measured stellar parameters that lie within our parameter range (see Eq.~\ref{eq:sp_cuts}). This gives us a total of 5\,222 spectra for 2\,053 solar twins. 
{The top panel of Fig.~\ref{fig:histograms} shows a histogram with the number of observations for the stars in the repeat observations sample. In general we have 2 or 3 observations for each star, about hundred stars with 3 and 4 observations and a few of stars having more than 4 observations.}

{For this sample we compute the uncertainties as follows: for each star we consider its observation of maximum SNR in CCD2, then for each repeat observation we compute the difference in the prediction compared to the one of maximum SNR. Finally, after doing it for all the stars in the sample, we divide the SNR range in bins of length 10 and we compute for each bin the standard deviation of the difference in predictions of the repeat observations within SNR bin to their respective maximum SNR prediction.} 

The bottom panel of Fig.~\ref{fig:histograms} shows a normalised density histogram of the repeat observation counts (in cyan) as function of the CCD2 SNR. In gray we show the observation counts for the entire solar twin catalogue.  The spectra of repeat observations cover well the SNR distribution of the entire catalogue. The figure shows that our sample is a good representation of the SNR of the entire GALAH~DR3 dataset.

\section{Impact of training choices on label recovery with The Cannon} \label{sec:results}



In this appendix section we discuss the results of different training choices on the label recovery of {\it The Cannon}. We perform the three main studies which are discussed in the following sections. The first study is the impact of the predicted labels and their uncertainties for different training set sizes. The second study is the reliability of the covariance uncertainties given the different training set sizes, and the third study is the accuracy and precision of the predicted labels when the training set has label outliers.

\subsection{Bias as function of training set size}

We consider only the high SNR sample of \highSNRsample\ stars described in Sect.~\ref{sec:data}. From there, we take 20\% for testing, and 80\% for training set selection. Thus, we have fixed 1\,008 stars for testing and 4\,032 stars to build different training sets. 

From the 4\,032 spectra available for training we study training sizes $N$ = 10, 20, 30, 50, 100, 150, 200, 250, 300, 350, 400, 500, 1\,000, 2\,000, 3\,000. We further choose randomly 10 subsamples of size $N_{1}=10$ and train 10 different {\it The Cannon} models, each one using a different subsample of size $N_{1}$ as its training set. This allows us to evaluate the impact of the choice of stars in each set. Then for $N_{2}=20$, we select the first subsample of size $N_{1}=10$ and randomly add the missing $N_{2}-N_{1}$ spectra adding up to a training set of $N_{2}$ spectra. This random addition of the first subsample of size $N_{1}$ is done 10 times, generating 10 subsamples of size $N_{2}=20$. We again train 10 different {\it The Cannon} models, each one using a different subsample of size $N_{2}$ as its training set. This procedure is followed for all the rest of the training sets and sizes.

\begin{figure}
    \centering
    \includegraphics[scale=0.25]{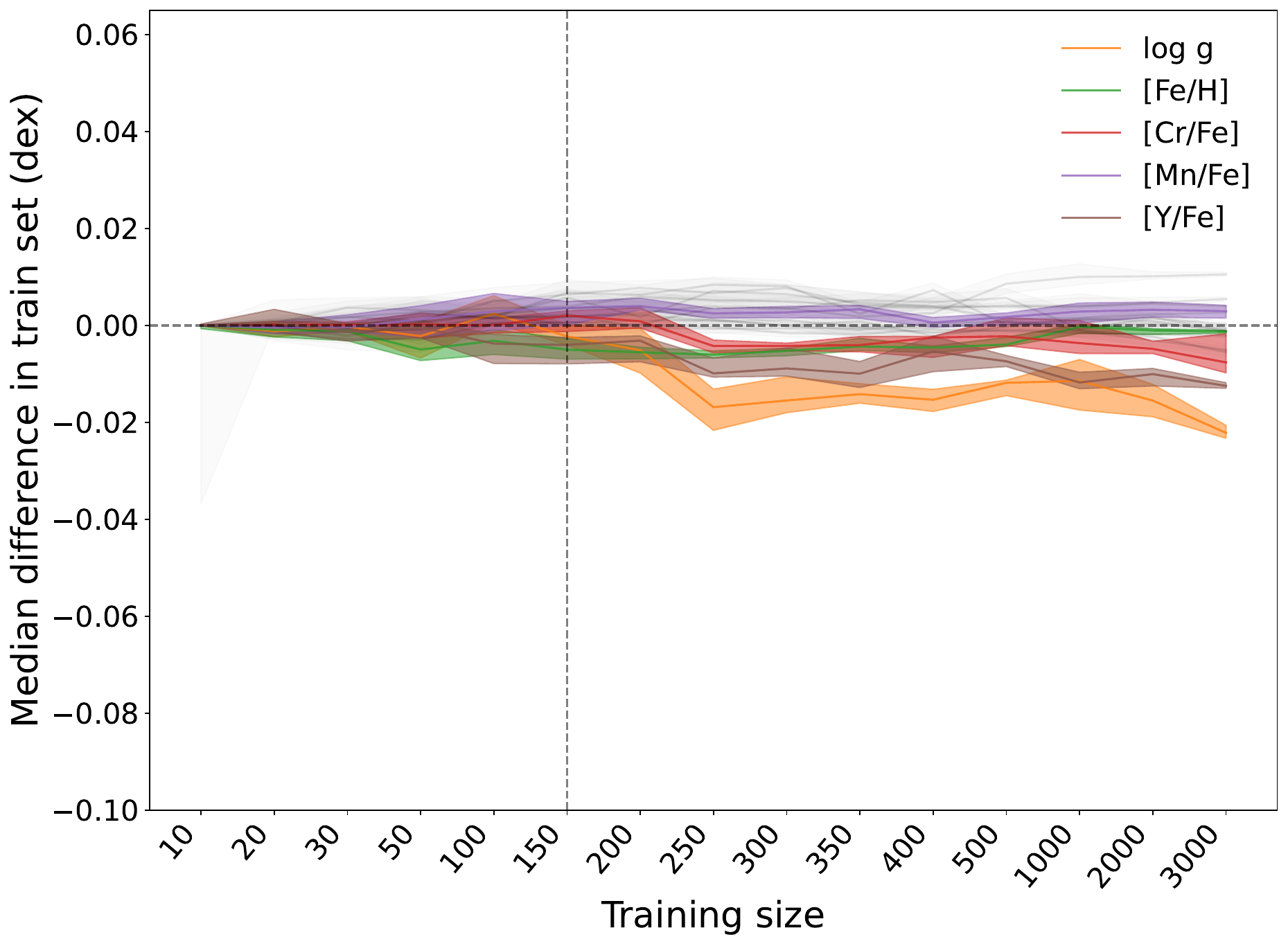}
    \includegraphics[scale=0.25]{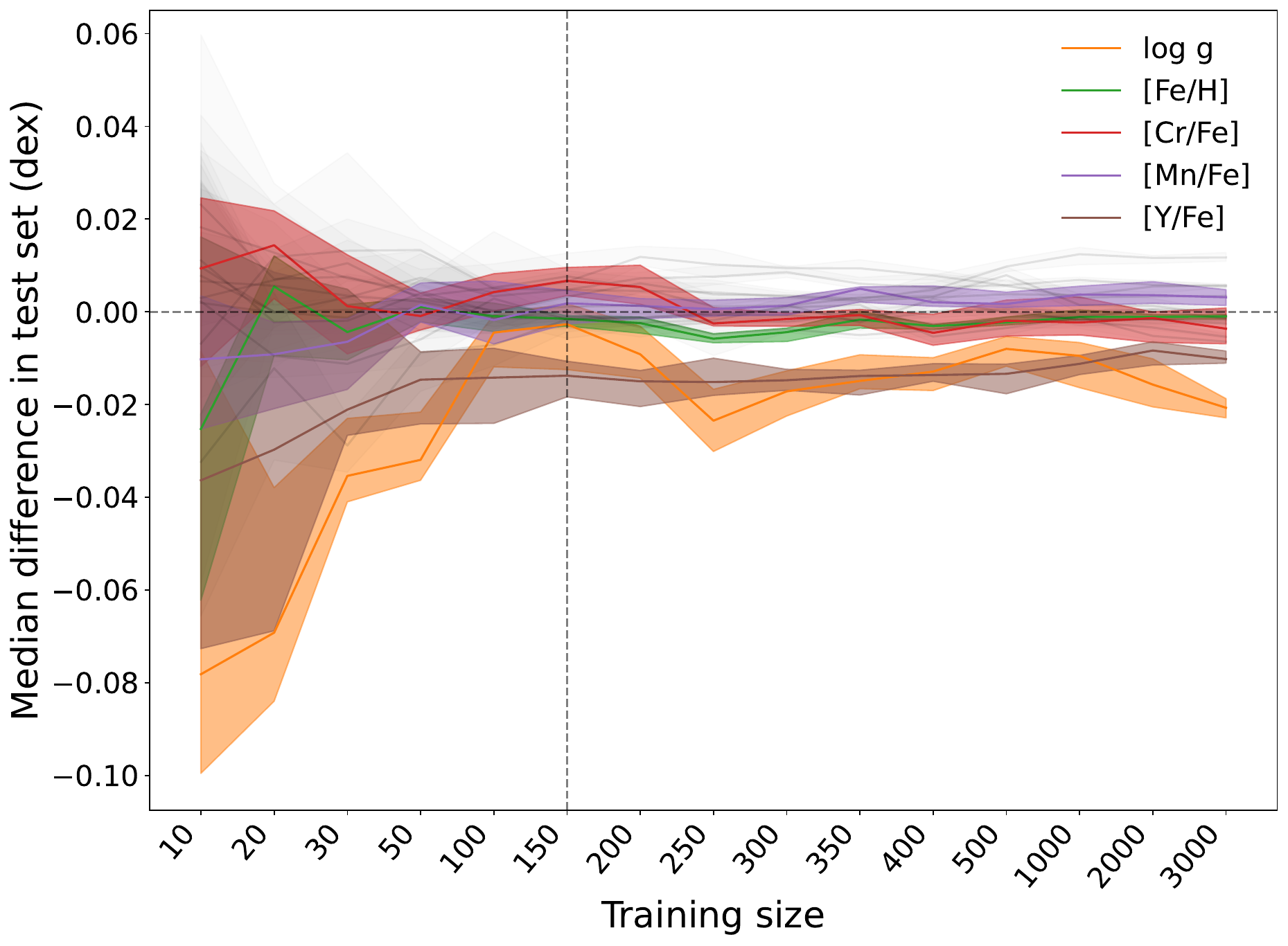}
    \caption{Top panel: Bias in train as function of training size $N$.  Bands represent the 16th and 84th percentile of the values obtained for the 10 different trained models with training sets of size $N$. Bottom panel: Bias Median in test as function of training size. For better visualisation of the results, effective temperature is not shown here. }
    \label{fig:trainsize_diff}
\end{figure}

\begin{figure}
    \centering
    \includegraphics[scale=0.25]{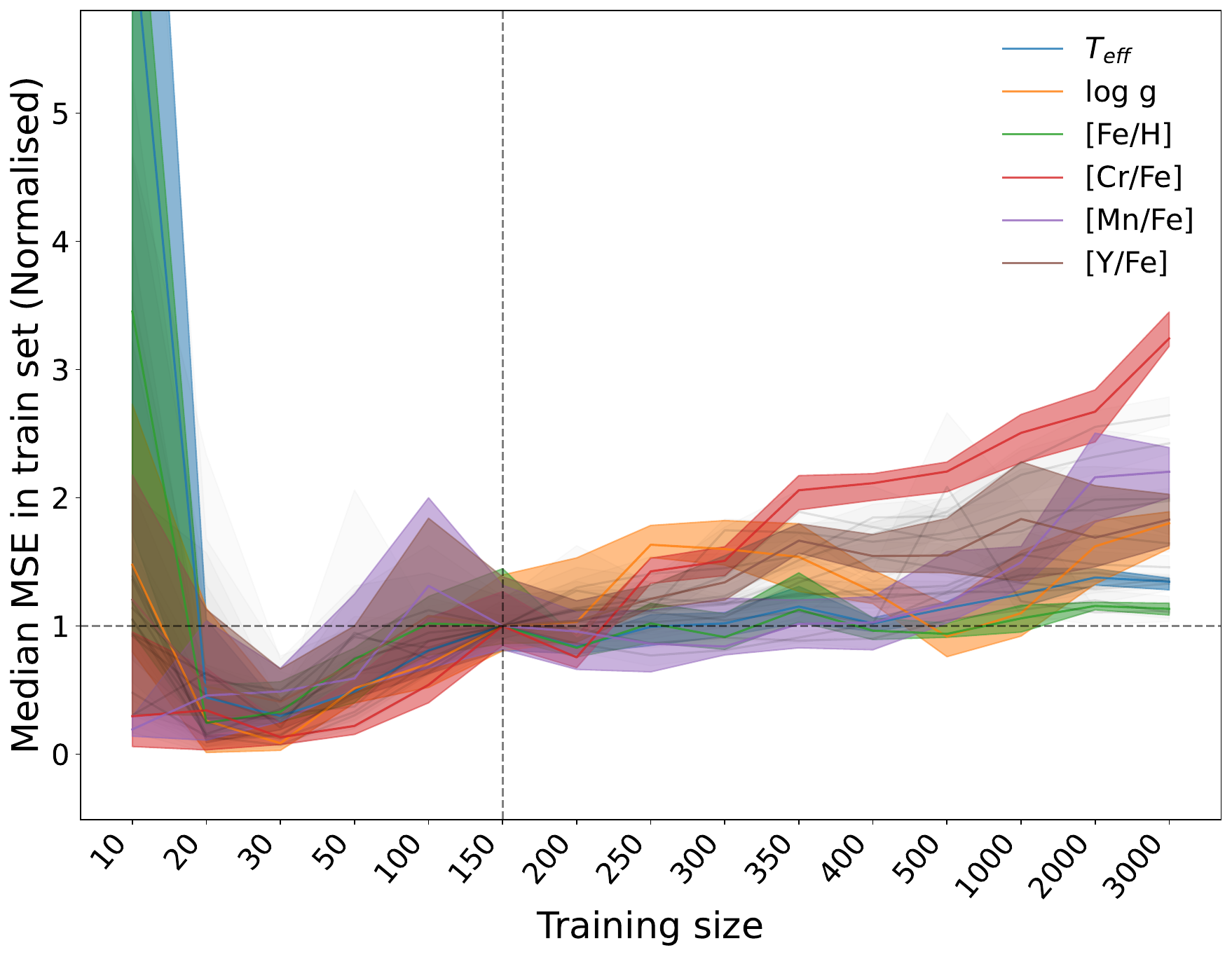}
    \includegraphics[scale=0.25]{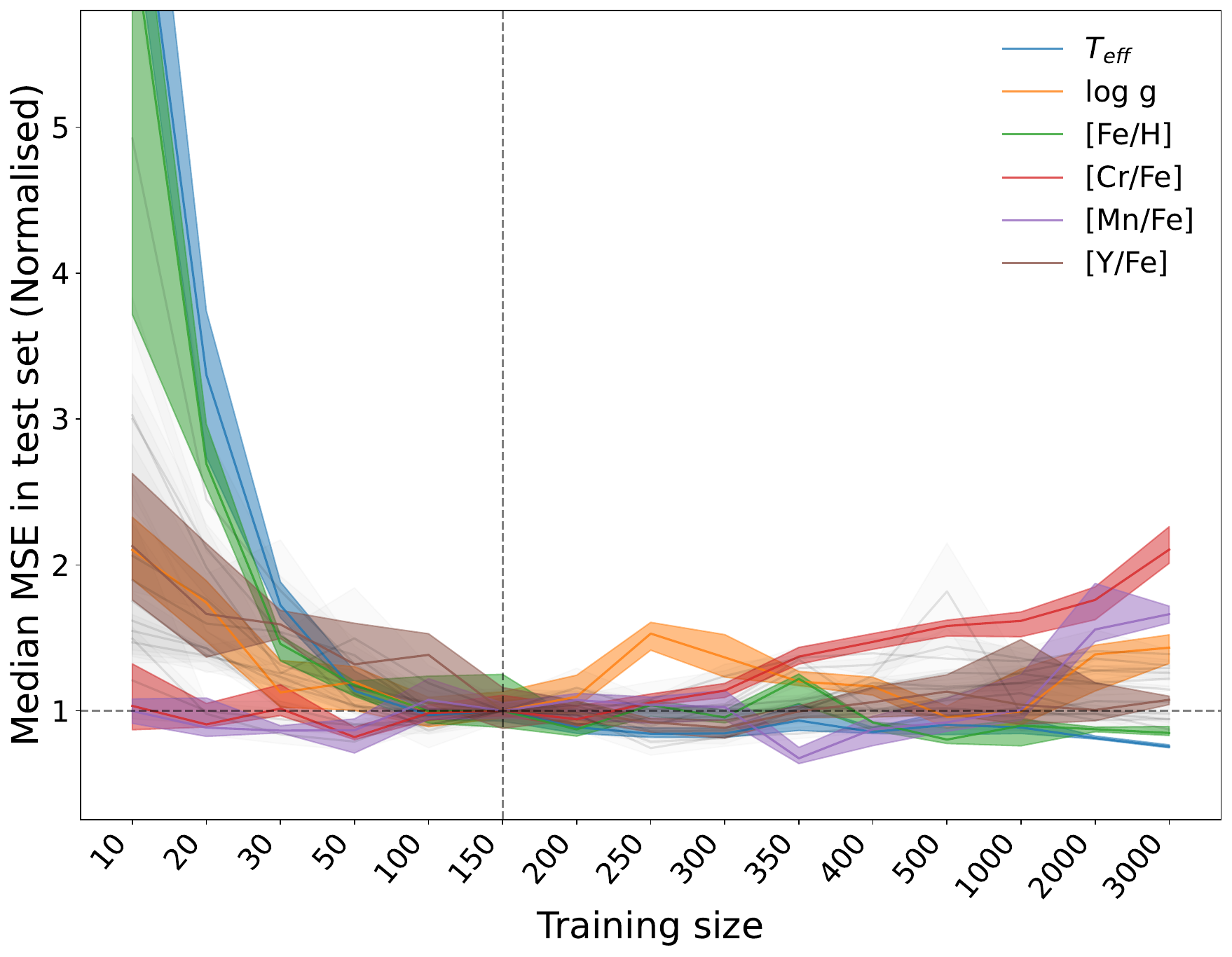}
    \caption{Top panel: Median of Mean Squared Error (MSE) in train  as function of training size $N$. Error bands represent the 16th and 84th percentile of the values obtained for the 10 different trained models with training sets of size $N$. Bottom panel: Median of MSE in test. Top and bottom panel normalised in value at training size $N=150$.}
    \label{fig:trainsize_mse}
\end{figure}

With these models, we predict the labels for the 1008 stars in our test set and compare them with the values in GALAH~DR3, which is shown in Fig.~\ref{fig:trainsize_diff}. In the top panel we show the median difference (bias) between the predicted labels and GALAH~DR3 in training as function of the training set size. Each line corresponds to a different label with error bands representing the 16th and 84th percentiles in the distribution of the results obtained by the 10 different \Cannon\ models. Some labels are highlighted with color as example. The bias between the predicted labels and GALAH~DR3 in test as function of training size is shown in the bottom panel. 
The difference between the GALAH~DR3 labels and predicted \Cannon\ labels are small for all training sizes $N$. \logg\ and [Y/Fe] show a minor bias increase with $N$ up to $0.01-0.02$ dex. For $N=10$, the differences are essentially zero for all the labels, showing the overfitting effect in \Cannon. 

We comment the bias in the parameter logg and its similarity with [Y/Fe], which are highlighted with orange and brown colors, respectively.   GALAH~DR3 surface gravities are not determined directly from the spectra, but from photometry and the parallax \citep{GALAH-DR3} because GALAH spectra do not contain sufficient dependency of this parameter. It is hence expected that the \Cannon\ model does a poor job in predicting this parameter. 

The impact of the bias in \logg\ with [Y/Fe] is  because the our \Cannon\ model considers the ionised Y lines in GALAH spectra. Ionised lines have a dependency on surface gravity. If the surface gravity is poorly determined, it is expected that a method deriving abundance of ionised lines of a given strength will respond by balancing the ill-determination of surface gravity with an ill-determination of that abundance. Most of the other abundances are derived from neutral lines, which are less sensitive to gravity.  From the bottom panel of Fig.~\ref{fig:trainsize_diff} we observe that for most of the labels the differences show a decreasing trend with training size $N$. The trend reaches a plateau at around $N=150$ which is marked with a vertical dashed line. 

To further assess the potential problems of overfitting in the training process, we compute the {\it Mean Squared Error (MSE)}. This allows us to assess the quality of the predictions made by the different \Cannon\ models in the test step, as well as the {\it self-test}, i.e. the test/label prediction step over the training set itself. In other words, we can evaluate how well the model recovers the labels for the stars considered in the training process. This is seen in Fig~\ref{fig:trainsize_mse}. The top and bottom panel show the median MSE in training and test, respectively, as a function of the training size $N$. Each line corresponds to a different label. We normalise the results to the ones obtained at $N=150$ for a better visualization, and the error bands cover the range of values between the 16th and 84th percentile over the distribution of the results obtained by the 10 different \Cannon\ models. 

When $N=10$ the MSE in train is large with a high discrepancy between the 10 different \Cannon\ models. 
The MSE significantly drops for $N=20$, but starts increasing again with the training size.  The bottom panel shows that the MSE in test starts high for small training sets, decreasing monothonically until a training size of $N=150$ and remain small for higher $N$. 
Our analysis shows that for a training set smaller than 150 stars, \Cannon\ is affected by overfitting if 14 labels from solar twin GALAH spectra are estimated. For training sets larger than 150 stars, we do not see a significant improvement in the bias and the MSE of the model. Therefore, we conclude that 150 stars is an optimal size for our \Cannon\ model of solar twin stars.

\subsection{Reliability of precision estimates (covariance uncertainties)}
\label{sect:reliability_uncertainties}

\begin{figure}
    \centering
    \includegraphics[scale=0.25]{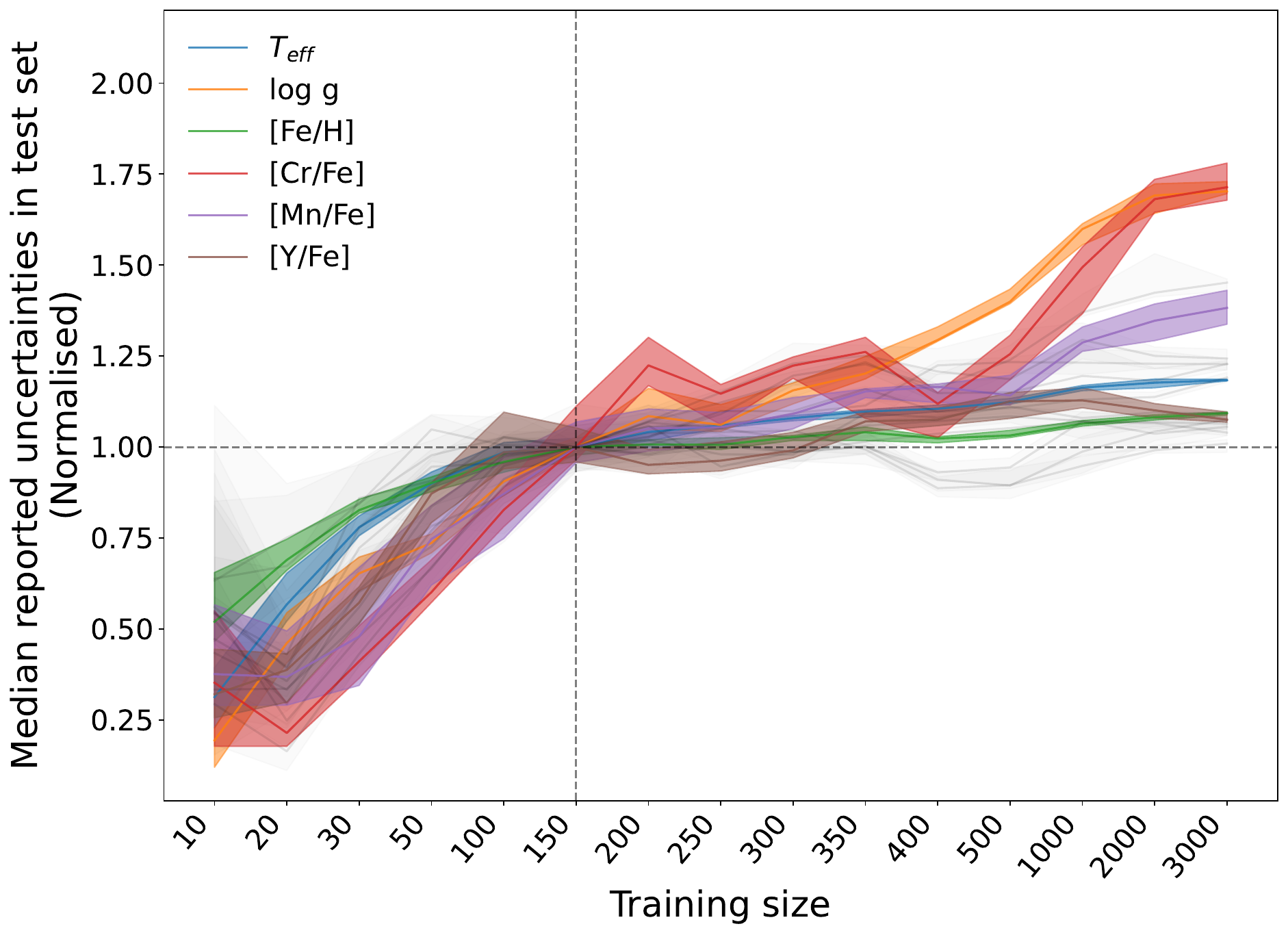}
    \caption{Median of the reported uncertainties in test as function of training size. Normalised at value in training size $N=150$. Error bands represent the 16th and 84th percentile of the values obtained for the 10 different trained models with training sets of size $N$.}
    \label{fig:median_test_uncertainties_norm150}
\end{figure}

{In Fig.~\ref{fig:median_test_uncertainties_norm150} we show the median reported internal uncertainties for \Cannon\ as a function of the training size $N$.  The lines and error bands follow the same definition as in Figs.~\ref{fig:trainsize_diff} and ~\ref{fig:trainsize_mse}. We recall that the internal uncertainties in \textsc{sme} are computed from the diagonal of the covariance matrix given by \textsc{sme} fitting procedures. With \Cannon\ we compute the internal uncertainties in the same way.}

There is a similar trend for all labels, where the median internal uncertainties in test increase with $N$. The stellar parameters \teff\ and \feh\ increase slowly, reaching a value up to 20\%  at $N=150$. When focusing on Y, we find that values of $150< N \leq300$ reports smaller uncertainties, but steadily increase for larger $N$. For \logg, Cr and Mn,  after $N=150$ the uncertainties considerably increase, reaching values nearly 70\% higher in the case of \logg\ and Cr, and 30\% higher for Mn. For the smallest training set sizes, we attribute the low uncertainties to overfitting. In particular for \logg\, we attribute this increase to the disagreement of spectroscopic and photometric \logg\ (compare to top panel of Fig.~\ref{fig:trainsize_diff}). For GALAH~DR3, \logg\ was actually not estimated from the spectra, but photometric information due to the usually lower information content in spectra. This approach can, however, fail for a variety of reasons like binarity or wrong photometric or astrometric information and thus introduce an increasingly systematic trend in the training. While we tried to exclude peculiar spectra from our training set via visual inspection, we suspect that for training sizes above 500, residual peculiar spectra caused the rise of training imperfections and the reported uncertainties. Some further imperfections are also not visible noted, and lead to training labels which are incorrect.  Such training imperfections, as evidenced by the rising MSE in Fig.~\ref{fig:trainsize_mse} for Cr, can then propagate into the test uncertainties.}

Hereafter we consider $N=150$ stars as our final training set size. We can conclude that this size is sufficient for making a {\it The Cannon} model that predicts most labels with high accuracy and sufficient  precision.  There is not much improvement in both accuracy and precision for higher values of $N$.

\subsection{Robustness against outliers}\label{sect:outliers}

\begin{figure}
    \centering
    \includegraphics[scale=0.28]{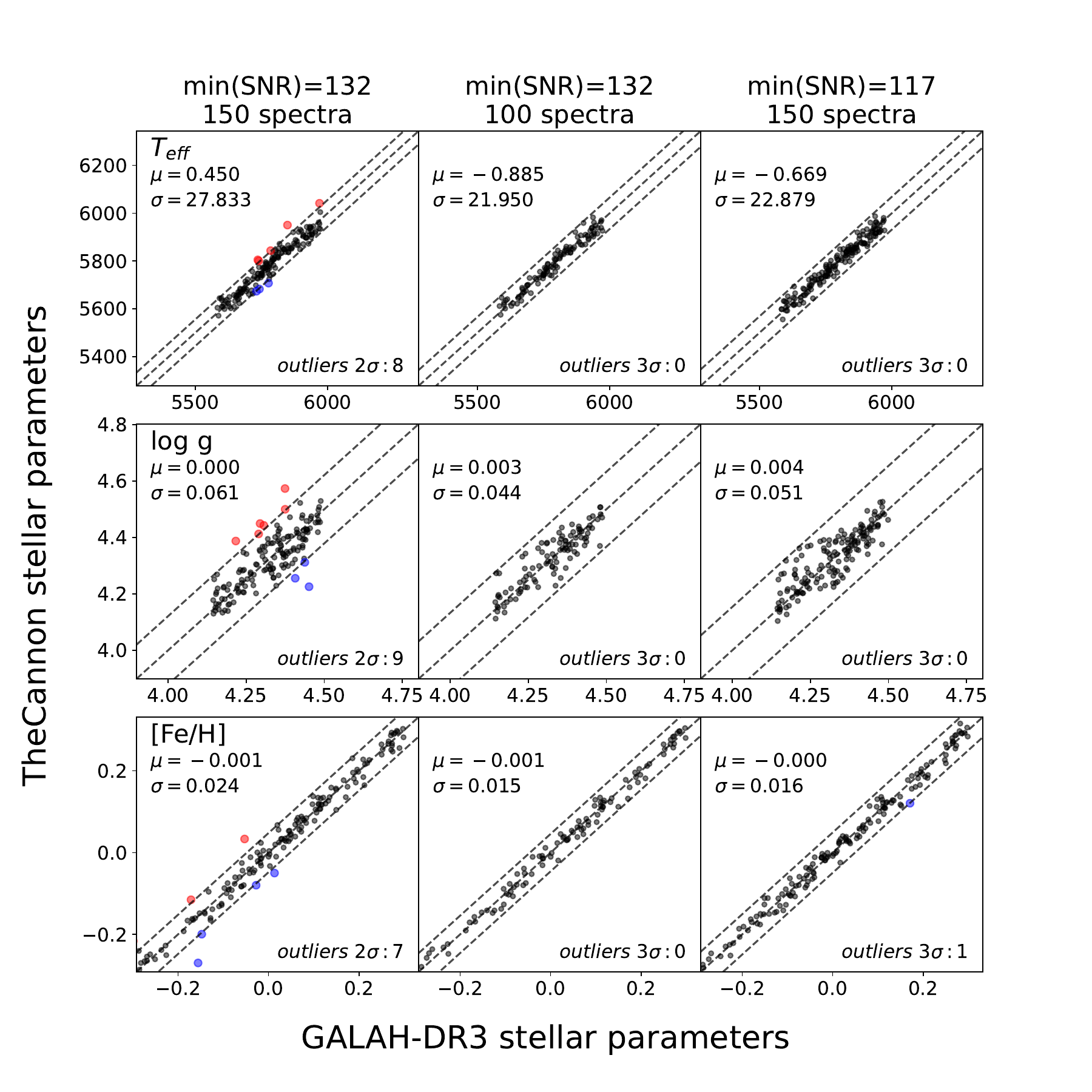}
    \caption{One-to-one comparison of stellar parameters \teff, \logg, \feh. Comparison of 3 setups/models. Model 1: 150 spectra of minimum SNR of 132 (left column), Model 2: 100 spectra of minimum SNR of 132 (middle column) and Model 3: 150 spectra of minimum SNR of 117 (right column) . In x axis, GALAH~DR3 labels. In y axis, \Cannon\ model estimates for labels. Outer dashed lines correspond to $2\sigma$ and $3\sigma$ boundaries for Model 1 and Models 2,3 respectively. Overestimates in red, underestimates in blue. Upper left and bottom right of each panel shows the median $\mu$ and standard deviation $\sigma$ of the difference, and the number of outliers found outside each boundary, respectively.} \label{fig:solar_twin_train_set_predict_1to1_3models_plot_16xfe_0}
\end{figure}

\begin{figure*}
    \centering
    \includegraphics[scale=0.28]{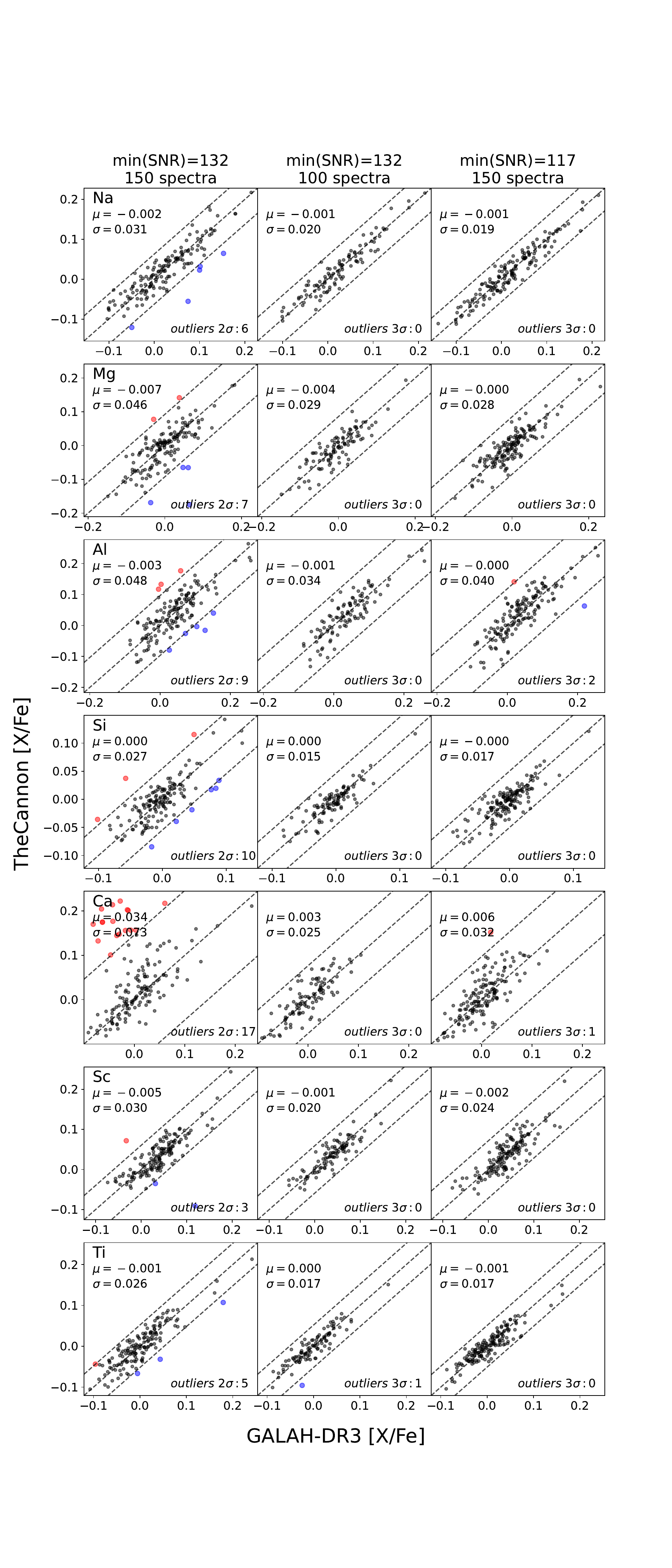}
    \includegraphics[scale=0.28]{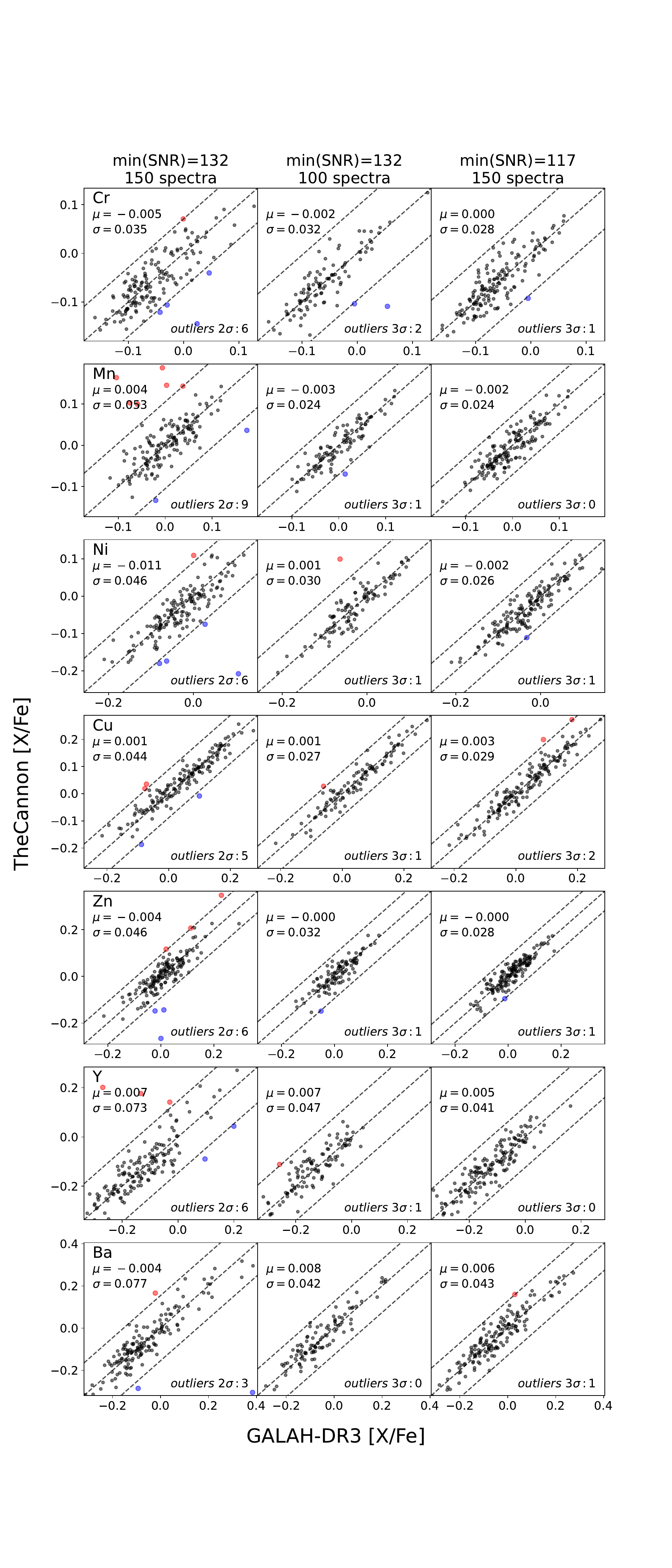}
    \caption{One-to-one comparisons for chemical abundances for the 3 setups/models. Model 1: 150 spectra of minimum SNR of 132 (left column), Model 2: 100 spectra of minimum SNR of 132 (middle column) and Model 3: 150 spectra of minimum SNR of 117 (right column) . In x axis, GALAH~DR3 labels. In y axis, \Cannon\ model estimates for labels. Outer dashed lines correspond to $2\sigma$ and $3\sigma$ boundaries for Model 1 and Models 2,3 respectively. Overestimates are plotted in red, underestimates are plotted in blue. Upper left and bottom right of each panel shows the median $\mu$ and standard deviation $\sigma$ of the difference, and the number of outliers found outside each boundary, respectively.} \label{fig:solar_twin_train_set_predict_1to1_3models_plot_16xfe_1_}
\end{figure*}

\begin{figure}
    \centering
    \includegraphics[scale=0.37]{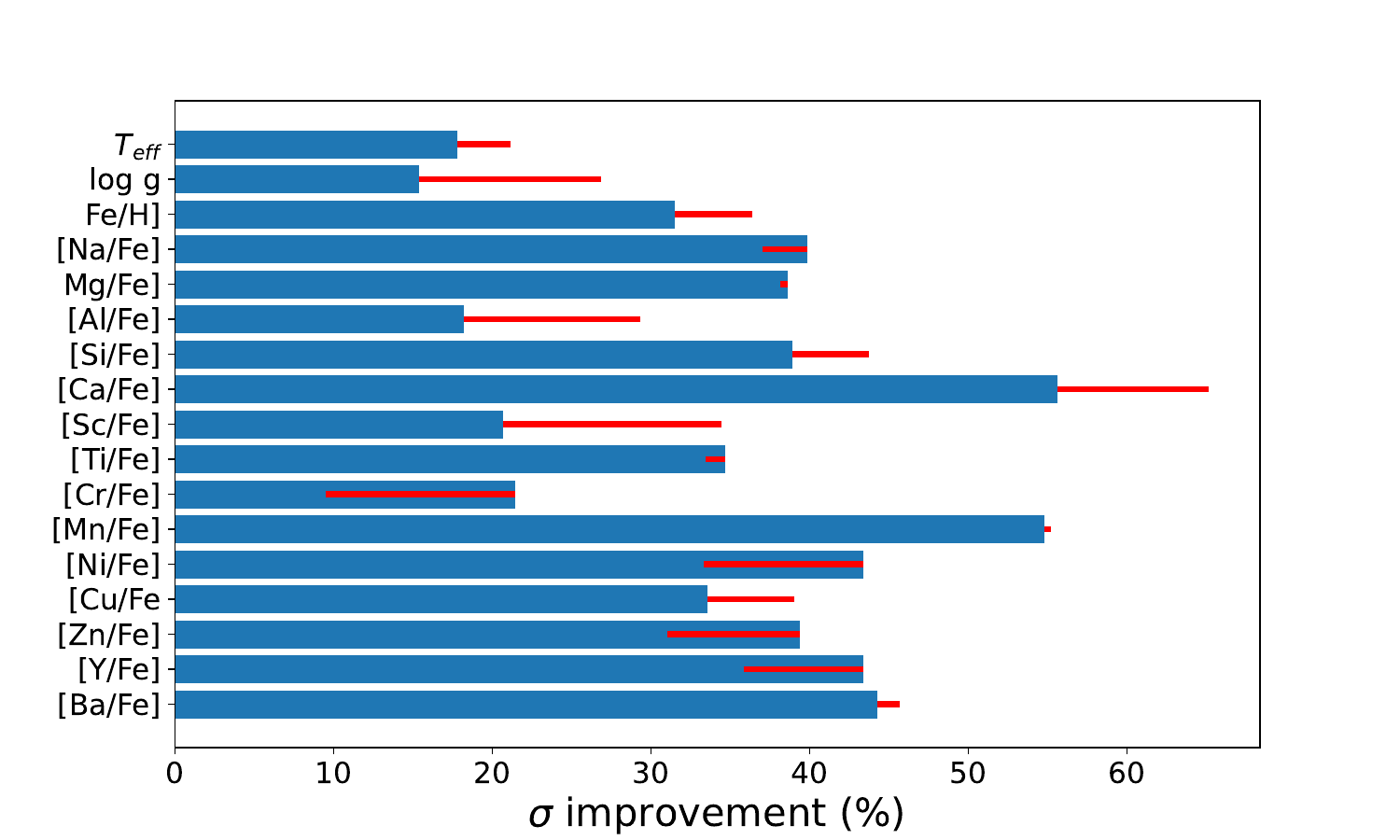}
    \caption{$\sigma$ improvements with respect to Model 1 for all labels (in percentage). Improvement given by model 3 in blue bins and improvement/decay in percentage made by model 2 in red lines.}
    \label{fig:sigma_improvement}
\end{figure}

We investigate how robust is our {\it The Cannon} model against potential outliers in the training set. We choose the optimal size of 150 stars for this analysis. To do so, we first select 150 stars with the highest SNR in our sample (hereafter Model 1). This yields a training set of stars with SNR above 132 in all CCDs. Then, we train a {\it The Cannon} model and perform a self-test. The results for Model 1 can be found on the left panels in Fig. \ref{fig:solar_twin_train_set_predict_1to1_3models_plot_16xfe_0} for stellar parameters  and Fig.~\ref{fig:solar_twin_train_set_predict_1to1_3models_plot_16xfe_1_} for chemical abundances. We set up the $3\sigma$ boundaries for all labels and define for each one of them the outliers those who lie outside this boundary. In the figures, these outliers are plotted with red and blue colours. In general the predictions about the parameters agree with GALAH~DR3. However, in Fig. \ref{fig:solar_twin_train_set_predict_1to1_3models_plot_16xfe_1_} we begin to see outliers for some of the chemical abundances, in particular as Ca and Mn. {There are four Ca lines, but two of them lie in telluric lines where the correction may not always be perfect \citep{galah-dr2}. In GALAH~DR3 the spectrum uncertainty is increased for these lines to account for strong blends of the telluric lines. Here with \Cannon\ we take the spectrum uncertainty directly from the database, obtaining overabundances of Ca because the model finds a high absorption feature (\citet{galah-dr2}).}

We took two different approaches to deal with the outliers.  The first one was to remove all the 50 outliers found in the self-test and train a new {\it The Cannon} model with the remaining 100 stars of minimum SNR 132 (hereafter Model 2). The second one was to remove the outliers in the first model and set a lower threshold in SNR to add more stars and build up a new training set of size 150 to then train a new {\it The Cannon} model. In this second case that this model produced new outliers, so we reiterated the process by removing such outliers and setting a lower minimum SNR threshold to build a new training set of size 150. After 4 iterations we converged to a {\it The Cannon} model trained with a set of 150 stars with minimum SNR of 117 (hereafter Model 3).

The results of these two approaches can be seen in the middle and right panels of Figs. \ref{fig:solar_twin_train_set_predict_1to1_3models_plot_16xfe_0} and \ref{fig:solar_twin_train_set_predict_1to1_3models_plot_16xfe_1_}, for Model 2 and Model 3, respectively. There are no outliers for the stellar parameters in both approaches. The same holds for the chemical abundances in Fig. \ref{fig:solar_twin_train_set_predict_1to1_3models_plot_16xfe_1_} where most labels have either no outliers or very few. 


\subsection{Choosing final training set} \label{sect:final_trainset}

From Figs. \ref{fig:solar_twin_train_set_predict_1to1_3models_plot_16xfe_0} and \ref{fig:solar_twin_train_set_predict_1to1_3models_plot_16xfe_1_} we can see that  Models 2 and 3 are an improvement with respect to Model 1 in terms of agreement in with respect to GALAH~DR3. 
Fig.~\ref{fig:sigma_improvement} shows the overall percentage improvement of the dispersion $\sigma$ in each label, given by the two latter models with respect to Model 1. The blue bins represent the percentage of improvement for Model 3, and the red lines represent the increment/decay in percentage of the improvement in $\sigma$ by the Model 2. For stellar parameters we observe a higher improvement given by Model 2, with a difference of 3.3\% , and 4.9\% for the \teff\ and metallicity,  respectively. For \logg\ we have a considerable difference of 11.5\% in favor of Model 2. For chemical abundances we observe negligible differences for  Mg, Ti, Mn, Ba where the differences are up to 1.4\%.  For  Al, Si, Ca, Sc, Cu we find higher differences up to 13.7\% in favor of Model 2, and for Na, Cr, Ni, Zn, Y we obtain differences up to 11.9\% in favor of Model 3. The mean improvement over all labels is 36.2\% and 34.8\% for models 2 and 3, respectively.


Taking into consideration that the overall difference of 2.6\% in mean improvement for all the labels is very small, we choose Model 3 for training. This model has an optimal size of 150 as well as a better coverage in the parameter space of stellar parameters and 14 chemical abundances for solar twins. 

\section{Stars selected for phylogenetic analysis} \label{sec:table_selected_stars}
Table~\ref{tab:selected_stars} shows the ages and eccentricities of 50 stars used in the tree. The information about the rest can be found online. 
\begin{table}
\begin{tabular}{|r|r|r|r|}
\hline
  \multicolumn{1}{|c|}{Tip ID} &
  \multicolumn{1}{c|}{Gaia DR3 ID} &
  \multicolumn{1}{c|}{Age (Gyr)} &
  \multicolumn{1}{c|}{$\epsilon$} \\
\hline
  0 & 5396076243592498944 & 8.74 & 0.6303\\
  1 & 3814762674970623104 & 7.50 & 0.6109\\
  2 & 6690566020667409024 & 8.49 & 0.5699\\
  3 & 56608356156072448 & 10.90 & 0.5324\\
  4 & 2532978746890377472 & 10.26 & 0.5305\\
  5 & 6557301912844603264 & 10.53 & 0.5257\\
  6 & 4710178345898217728 & 10.76 & 0.5183\\
  7 & 3747102371529922560 & 10.11 & 0.5142\\
  8 & 2582225773914697728 & 5.00 & 0.5024\\
  9 & 6247059932386652288 & 3.91 & 0.4902\\
  10 & 6483285057806757376 & 9.39 & 0.4796\\
  11 & 4475030814892134912 & 11.21 & 0.4793\\
  12 & 5416434491653412480 & 9.05 & 0.4750\\
  13 & 3201058165300460928 & 9.51 & 0.4748\\
  14 & 3910048792175067648 & 10.14 & 0.4660\\
  15 & 5827969736762248192 & 5.17 & 0.4607\\
  16 & 6665136481380692992 & 5.97 & 0.4605\\
  17 & 6811053318039677184 & 10.53 & 0.4584\\
  18 & 6094499086259119360 & 9.81 & 0.4574\\
  19 & 2558440382468221184 & 10.33 & 0.4530\\
  20 & 5565471677892685184 & 10.29 & 0.4527\\
  21 & 3206464940714422144 & 7.26 & 0.4523\\
  22 & 3462833192176962688 & 8.99 & 0.4523\\
  23 & 3753282382791867904 & 9.78 & 0.4504\\
  24 & 6579941746317849344 & 5.57 & 0.4498\\
  25 & 3203847042185006080 & 11.39 & 0.4497\\
  26 & 2616281875274482560 & 9.26 & 0.4491\\
  27 & 685943836661554944 & 7.10 & 0.4469\\
  28 & 5566658772490401920 & 8.59 & 0.4453\\
  29 & 6687193131310348032 & 11.20 & 0.4448\\
  30 & 5780117135278458112 & 10.51 & 0.4446\\
  31 & 3857760795161492992 & 10.32 & 0.4433\\
  32 & 3405317601488435328 & 10.27 & 0.4410\\
  33 & 5484461722739849984 & 9.42 & 0.4406\\
  34 & 6583876554835641088 & 4.42 & 0.4405\\
  35 & 6140741109344938368 & 10.20 & 0.4404\\
  36 & 3393159923462202624 & 7.76 & 0.4404\\
  37 & 3694036023363825152 & 8.24 & 0.4401\\
  38 & 3618875671935767936 & 10.92 & 0.4401\\
  39 & 6033793262610025984 & 7.65 & 0.4314\\
  40 & 2600019239306534528 & 9.16 & 0.4310\\
  41 & 2715831692413101056 & 8.16 & 0.4301\\
  42 & 6147111267399070464 & 10.34 & 0.4301\\
  43 & 6379802489437427840 & 9.64 & 0.4297\\
  44 & 6456328640466199424 & 10.45 & 0.4292\\
  45 & 6144956155890048256 & 9.97 & 0.4289\\
  46 & 4823564043700716672 & 8.10 & 0.4282\\
  47 & 5722776126416465792 & 10.66 & 0.4274\\
  48 & 5582111338970433664 & 11.23 & 0.4265\\
  49 & 5451886835341219712 & 11.44 & 0.4260\\
  50 & 6656031455656664192 & 8.57 & 0.4257\\
  \hline
\end{tabular}
\caption{Example of 50 stars selected for phylogenetic analysis (see Fig.~\ref{fig:mcctrees_ecc}). Tip ID corresponds to the integer displayed in the tips of the trees. For reference, we show the Gaia DR3 IDs, and the ages and eccentricities ($\epsilon$). The information of the rest of the stars can be found online.}\label{tab:selected_stars}
\end{table}

\bsp	
\label{lastpage}
}
\end{document}